\documentclass[opre,nonblindrev]{informs4}
\usepackage[utf8]{inputenc}
\usepackage{eqndefns-left}
\Equationvalidatefalse
\RequirePackage{tgtermes}
\RequirePackage{newtxtext}
\RequirePackage{newtxmath}
\RequirePackage{bm}
\RequirePackage{endnotes}
\usepackage[normalem]{ulem}

\OneAndAHalfSpacedXI

\usepackage{algorithm}
\usepackage{algpseudocode}
\usepackage{float}
\usepackage{afterpage}
\usepackage{tikz}
\usepackage[hidelinks]{hyperref}
\usepackage{booktabs}
\usepackage{array}
\usepackage{graphicx}
\usepackage{subcaption}
\usepackage{mdframed}
\usepackage{multirow}
\usepackage{enumitem}

\newcounter{marginparcounter}

\usepackage{natbib}
 \bibpunct[, ]{(}{)}{,}{a}{}{,}%
 \def\bibfont{\small}%
\newcolumntype{L}[1]{>{\raggedright\let\newline\\\arraybackslash\hspace{0pt}}m{#1}}

\EquationsNumberedThrough
\TheoremsNumberedThrough
\ECRepeatTheorems
\newtheorem{result}{Result}

\MANUSCRIPTNO{}

\begin{document}

\RUNAUTHOR{Hathaway, Kagan, and Jones}
\RUNTITLE{The Personal Productivity Bias in Emergency Department Patient Assignment}
\ECRUNAUTHOR{}
\ECAUpunct{}
\TITLE{One at a Time? The Personal Productivity Bias in Emergency Department Patient Assignment}

\ARTICLEAUTHORS{%
\AUTHOR{Brett A.\ Hathaway\textsuperscript{a}, Evgeny Kagan\textsuperscript{b}, John R. Jones\textsuperscript{c}}
\AFF{\textsuperscript{a}Marriott School of Business, Brigham Young University, \EMAIL{brett\_hathaway@byu.edu}\\
\textsuperscript{b}Carey Business School, Johns Hopkins University, \EMAIL{ekagan@jhu.edu}\\
\textsuperscript{c}UnityPoint Health Sioux City, \EMAIL{john.jones@unitypoint.org}}
} 

\ABSTRACT{%
Emergency departments (EDs) often use a shared-queue setup in which physicians self-assign cases from a pool of triaged patients. We conduct a multi-method study to examine this self-assignment behavior and its effects on system performance. Using data from five EDs spanning 1.4 million patient visits, we show that \textit{batching}, i.e., self-assigning multiple patients at once, is common and associated with longer stays for batched patients, even after controlling for clinical acuity, physician fixed effects, and ED congestion. We then develop a continuous-time queueing model that characterizes the optimal self-assignment policy under individual and group throughput incentives. We use the model predictions to test experimentally with 203 healthcare workers and 73 ED physicians whether batching is a rational response to incentives or a deeper behavioral tendency that persists independent of incentives. Indeed, batching is pervasive across both samples, with 94\% of healthcare workers and 73\% of physicians choosing to batch even when it reduces their own payoffs---a behavior that we term \textit{the personal productivity bias}. Together, these results suggest that compensation redesign alone is unlikely to eliminate batching, and suggest changes to the assignment interface in the electronic health record system as a more promising remedy.}


\maketitle
\vspace{-0.07cm}

\section{Introduction}\label{sec:Intro}
Emergency departments in the United States treat over 150 million patients annually \citep{CDCFastStats2022}, and even small inefficiencies in patient flow can translate into increases in length of stay and costs \citep{hoot2008systematic,morley2018emergency}. A potentially important source of such inefficiency is how patients are assigned to physicians. Most EDs use a real-time tracking board maintained by the electronic health record system that displays triaged patients, and physicians can claim unassigned cases before beginning treatment. Many physicians \textit{batch}, i.e., reserve multiple cases at once. 
This paper uses a combination of archival data analysis, stochastic modeling and behavioral experiments to better understand batch-assignment behavior and to examine its consequences for system performance.

Physicians may batch for several reasons. Batching guarantees an uninterrupted personal workload (at least in the short term) and reduces exposure to idle time between patients. This can be appealing financially, as physicians' compensation packages often include a piece-rate component, and also psychologically, as physicians may prefer action over inaction and may be averse to idleness or administrative work. Some physicians may genuinely believe that batching is good for the system. Claiming several cases at once may feel like an efficient way to move patients forward in the healthcare delivery process and reduce the backlog. If patient assignment incurs a fixed setup cost (switching from patient treatment to administrative tasks), batch-assigning multiple patients may indeed be more efficient. At the same time, a physician may fail to see that a claimed but unseen patient must wait for the physician to work through other assigned cases, and is simultaneously unavailable to colleagues with idle capacity.  

Despite its importance, physician-to-patient assignment-batching has received almost no empirical attention; the only prior study we are aware of is \citet{imhoff2022batched}, a single-site, correlational analysis of assignment batch size and patient length of stay. The goals of our paper are therefore (i) to document the prevalence of assignment batching across several urban and suburban EDs with different patient mixes over a multi-year period, (ii) to understand when and why physicians batch, with a particular focus on whether this behavior reflects economic incentives or a deeper behavioral tendency (``bias''), and (iii) to use these insights to test recommendations for redesigning incentive systems, patient assignment processes and decision-support interfaces.

\subsection{Research Design and Preview of Results}
We address the above questions using a multi-method approach that combines micro-level case data from five emergency departments (over 1.4 million visits spanning nearly ten years), an analytical model of physician self-assignment in a shared-resource queueing system, and pre-registered behavioral experiments with healthcare workers and practicing emergency physicians. The field data document the prevalence of batching, its variation across physicians and system conditions, and the relationship between batching and length of stay. The analytical model formulates self-assignment as a continuous-time Markov chain with stochastic arrivals and treatment times and characterizes optimal assignment policies under different types of incentives. The experiments test the model predictions with respect to whether batching persists under different compensation structures, and whether informational interventions can reduce it.

In the field data, we first document that batching is widespread but varies substantially across individual physicians, with some almost never batching while others do it frequently. Batching behavior is also load-adaptive: physicians are more likely to batch when their current caseload is low and when many unassigned patients are available. That is, physicians use batching as a strategy to manage their personal workload. We then examine the relationship between batching and patient outcomes. After controlling for physician fixed effects, clinical acuity, patient demographics, temporal patterns, and ED congestion, batching is associated with statistically and practically significant increases in the length of stay of batch-assigned patients. This finding is consistent across all five hospitals despite differences in patient populations, staffing models, and compensation structures, with average effects ranging from approximately 9\% to 12\%.

The prevalence of batching and its effect on patient stays raise a natural question: why do physicians batch? Are they responding rationally to financial incentives? Or, is batching a behavioral tendency that would persist even if such incentives were removed? To formalize this, we develop an analytical model of physician self-assignment. The model represents the ED as a continuous-time Markov chain with stochastic arrivals and treatment times, where multiple physicians share a finite pool of patients. We first analyze the problem from the perspective of an individual physician maximizing personal throughput. Indeed, under piece-rate compensation, batching is  optimal because it reduces the physician's exposure to stochastic idle time between cases. Under group throughput incentives, however, assigning exactly one patient at each decision epoch maximizes system throughput and minimizes expected patient sojourn time. The intuition is that any policy assigning more than one patient weakly delays service initiation for at least one patient, and these delays propagate to later completion times and higher congestion. 

We use the model predictions as theoretical benchmarks for controlled human-subject experiments. In the first experiment ($N=203$), we recruit healthcare workers on Prolific to make self-assignment decisions under different compensation schemes that vary whether batching or not batching is the payoff-maximizing strategy. Importantly, we hold incentive strength (payoff improvement from following the optimal over the suboptimal policy) constant across all conditions. Under individual throughput incentives, where batching is optimal, 94\% of participants indeed batch. Surprisingly, however, under group throughput incentives, where not batching is optimal, 94\% of participants continue to batch. That is, decision-makers continue to batch despite it being bad for their payoffs. Reframing the incentives in terms of patient sojourn time rather than group throughput does not reduce batching. However, a one-sentence informational nudge that highlights the consequence of batching for partner idleness significantly reduces batching (by 17 percentage points).

In the second experiment ($N=73$), we use a verified professional community to recruit practicing emergency physicians. We assign them to two treatments that are most relevant to the research question: group throughput incentives with and without the informational nudge. Overall, while physicians batch at lower rates than the Prolific sample, the majority still continue to do so. Under group throughput incentives, 73\% of physicians choose to batch despite it being the payoff-dominated strategy. The nudge, again, reduces batching by 17 percentage points. Post-experiment surveys are consistent with a behavioral rather than an economic explanation. Physicians who batched report wanting to stay busy and avoid idle time and describe batching as an ingrained professional habit. We conclude by discussing the implications of these results for models of optimal incentive and queue design (pooled vs. dedicated) in service operations.

\subsection{Contributions}
We make three contributions. First, we provide the first large-scale empirical documentation of assignment batching in emergency departments and show that it is associated with an increase of approximately 9\% to 12\% in length of stay for batched patients, relative to single-assignment patients seen by the same physician under similar conditions. While batching behavior has been studied in the operations literature \citep[e.g.,][]{feizi2023batch}, prior research has focused on \textit{admission} batching, i.e., delaying hospital admission orders for patients who have already been treated in the ED. In that setting, batching allows physicians to consolidate administrative tasks and is associated with higher individual productivity. In contrast, we study \textit{assignment} batching, i.e., the upstream decision of how many patients to claim from the shared pool before initiating care. In our setting, there are minimal efficiency gains for the physician since self-assigning a patient is done through a single click. Nevertheless, consistent with \cite{feizi2023batch} we also find that batching behavior may degrade system performance, and we identify the behavioral mechanisms that drive it.

Our second contribution is methodological. To set up theoretical benchmarks for rational behavior, we develop a continuous-time Markov chain model of patient-physician assignments and prove that one-at-a-time assignment is optimal for patient sojourn time and system throughput. A key feature of the model is that it is designed to be testable in the laboratory while preserving model elements that make the problem realistic (random arrivals and service times, shared capacity, endogenous congestion, etc.). Testing models of this type in an experiment presents two known challenges. First, policy performance in sequential decision problems tends to be insensitive to deviations from optimality especially within the short horizon of a lab session \citep{seale1997sequential,kagan2025sequential}. This makes it difficult to generate payoff differences that would make it worthwhile for a decision-maker to try to find the optimal solution. We address this by finding  system parameters that make the (expected and realized) payoff gap between optimal and suboptimal strategies non-trivial (in size) and constant across treatments. Second, standard approaches to testing behavior in sequential decision tasks include using random termination to represent discounting in infinite-horizon settings \citep[e.g.,][]{dal2019,RosokhaWei2024} or excluding warm-up and/or end-of-horizon decisions to focus on steady-state behavior \citep[e.g.,][]{kim2020admission,kim2024admission}. Our approach avoids this by building on the finite-horizon framework in \citet{HathawayKaganDada2023}, and formulating easily interpretable terminal conditions that induce stationary behavior (in the optimum) throughout the decision horizon. In particular, our approach extends \citet{HathawayKaganDada2023} from a discrete-time setting with deterministic times to a continuous-time setting with stochastic arrivals and service times.\footnote{A further difference is that  \citet{HathawayKaganDada2023} use early termination with continuation payoffs when decisions would extend beyond the horizon. In contrast,  we allow the system to run through the full horizon with participants compensated based on the state at the end.}

Third, we identify a robust behavioral bias in physician self-assignment decisions. A growing experimental literature documents deviations from optimal policies in healthcare and service operations, including admission control biases \citep{kim2020admission, kim2024admission}, state-dependent under- and over-testing \citep{kremer2023mismanaging}, biased response to different compensation systems in customer service \citep{HathawayKaganDada2023}, and suboptimal provision of effort in shared queues \citep{shunko2018,RosokhaWei2024}. We contribute to this literature by identifying the \textit{personal productivity bias} -- a preference for the strategy that keeps the decision-maker busy (at least in the short term) over the strategy that maximizes individual and system performance (in the long term). Unlike the biases documented in prior work, which are caused by miscalibrated responses to system state, complexity of incentives, or strategic behavior, the personal productivity bias persists when the optimal policy is not state-dependent, and incentives are transparent and independent of the behavior of others.

\section{Literature}\label{sec:Literature}
Our work builds on and contributes to three streams of literature: empirical work on physician workflow and ED operations, stochastic models of multi-server service systems, and behavioral experiments in service and healthcare operations. 

\subsection{Empirical Studies of Physician Workflow and ED Operations}\label{sec:lit:empirical}
A large empirical literature in healthcare operations uses administrative data to study how physicians manage workload, sequence tasks, and respond to congestion. A first stream documents \emph{load-adaptive} behavior at the patient level: physicians speed up when the system is busy \citep{KCTerwiesch2009,batt2017early}, discharge patients earlier from intensive care \citep{KCTerwiesch2012}, and may even let quality erode \citep{BerryJaekerTucker2017}. \citet{TanNetessine2014} document analogous workload effects in a service (restaurant) context. Our field data show a related and previously undocumented behavior: rather than working faster or slower in response to congestion, physicians adjust \textit{how many} cases they claim from the shared pool. In contrast to most load-adaptive responses, which tend to improve short-run flow at some cost to quality, assignment batching is associated with longer length of stay across all five EDs in our sample.

A second stream in this literature examines the consequences of physician discretion over which cases to take and in what order. \citet{IbanezClarkHuckmanStaats2018} show that radiologists deviate from FIFO and that these task-ordering choices have measurable productivity consequences; \citet{FreemanSavvaScholtes2017} study maternity units and show that workload affects midwives' referral behavior. Closer to our setting, \citet{KCStaatsKouchakiGino2020} document a \textit{task completion preference} among ED physicians: under high workload, physicians select easier cases, which increases their short-term throughput but decreases complexity-adjusted productivity. We will also examine load-adaptive behavior, but will be focusing on assignment batching rather than on cherry-picking behavior. To our knowledge, \citet{KCStaatsKouchakiGino2020} is the only prior paper to combine field data from emergency departments with a laboratory experiment to understand the mechanism driving the effect in the field, and we adopt a similar multi-method approach here. 

\citet{ChanGentzkowYu2022} use administrative data to identify physician-level variation in diagnostic style. \citet{SongTuckerMurrell2015} find that ED patients have longer stays under pooled than under dedicated queues, which they attribute to weaker physician ownership of cases drawn from a shared pool. The closest to us in this stream is \citet{feizi2023batch}, who study \emph{admission} batching (bundling of post-treatment admission orders) and show that batching helps physicians work more efficiently by consolidating administrative work. Our setting is a mirror image of \cite{feizi2023batch}. Unlike patient admission, which requires significant paperwork and takes a non-trivial amount of time, patient assignment takes only a single click, so there are no economies of scale that would rationalize batching it. 

The only prior study of assignment batching that we are aware of is \citet{imhoff2022batched}, a three-month retrospective analysis of resident physicians at a single urban academic ED during the COVID-19 pandemic. Their findings provide initial evidence that assignment batching is common and is associated with longer length of stay. However, the large presence of COVID-19 cases in their patient data (and potential idiosyncrasies that this may cause) makes it difficult to reach definitive conclusions. We generalize their initial findings to five urban and suburban EDs and 1.4 million visits spanning nearly a decade, study attending physicians whose compensation includes a productivity-linked component, and add physician fixed effects to absorb time-invariant physician characteristics and better identify the focal effect.

\subsection{Models of Shared-Pool Service Systems}\label{sec:lit:models}
Our analytical model is related to a queueing-control literature on multi-server systems with a shared pool of jobs. The bulk of that work studies systems in which a central dispatcher allocates jobs to parallel servers and asks when pooling outperforms dedicated assignment with respect to sojourn time or throughput \citep{MandelbaumReiman1998Pooling, Whitt1999Partitioning, vanDijkvanderSluis2009Pooling}. Within healthcare, \citet{SaghafianStreaming2012} show that routing patients into separate fast-track and main streams (rather than a single shared queue) can shorten time to provider for low-acuity patients.  \citet{ArgonZiya2009Priority} study how a server should prioritize across customer types when their identities are revealed through imperfect signals; and \citet{HoppIravaniYuen2007} examine service systems in which workers exercise discretion over how much effort to put into each task. A related line of work studies contract design in service queues, where the principal chooses a compensation structure that aligns server behavior with system goals \citep{GilbertWeng1998Incentive, CachonZhang2007FastService, ShumskyPinker2003}. We return to this literature in \textsection\ref{subsec:DiscussionModels}, where we discuss how our findings refine the standard monetary-utility framing. What sets our model apart is that routing is performed by the servers themselves, and each server can pull multiple jobs from the pool at a single decision epoch. Pulling extra jobs may benefit the individual server but imposes an externality on the system, which we characterize analytically and then test in the laboratory.

\subsection{Behavioral Experiments on Service Decisions}
A growing body of experimental work studies individual decision-making in service environments with stochastic arrivals, stochastic service times, and congestion. Closest to our work is \cite{kim2020admission} who examine admission control in an ICU setting and find that decision-makers have an occupancy-information bias: they under-admit patients when occupancy is visible, even when the optimal threshold policy calls for admission. \cite{kim2024admission} extend this to settings with diagnostic uncertainty. \cite{kremer2023mismanaging} study a related problem in which a server must decide when to stop testing and commit to a diagnosis under congestion pressure; they find that subjects over-test when the queue is short and under-test when it is long. \cite{HathawayKaganDada2023} study a gatekeeper who must decide whether to keep or transfer a customer, trading off the number of customers served against spending time with the current customer; they show that decision-makers overreact to transfer cost. Closest to us are \cite{RosokhaWei2024} who study effort provision in a shared queue and find that workers under-provide effort relative to the cooperative optimum, though cooperation improves when the game horizon is longer. Across these studies, a common finding is that human decision-makers deviate systematically from optimal policies, and that these deviations depend on the structure of information, incentives, and the service environment.

Our paper contributes to this literature in two ways. First, we study a decision that has not been previously examined: how many cases to self-assign from a shared pool. Unlike admission control (accept/reject) or effort provision (how hard to work), the batching decision involves choosing the \textit{size} of one's workload. Second, while prior work documents deviations from optimality, we identify a specific behavioral mechanism (personal productivity bias), whereby decision-makers prefer strategies that increase their own workload (in the short-term) but may reduce individual and system performance (in the long-term). This bias is distinct from the congestion neglect, anchoring, or free-riding documented in prior studies.


\section{Field Data}\label{sec:FieldData}
We use field data to show that batching is widespread in practice and associated with longer length of stay for batched patients across all five emergency departments in our sample. The field data, however, do not speak to \textit{why} physicians batch and what can be done about it -- questions that we will address in \textsection4-5 by developing an analytical model and testing it in  controlled experiments.

\subsection{Setting and Data}

\subsubsection{Environment}

Our empirical setting consists of five emergency departments (EDs) within a regional hospital network in the United States, with data spanning January 2013 to April 2022. Each ED is staffed by an independent physician group whose members are not hospital employees but instead contract with the hospital network to provide emergency care. A coauthor of this study was president of one of these groups and worked to obtain access to the administrative and clinical data across all five EDs in the network.\footnote{This study was reviewed and deemed exempt by the  Institutional Review Board of the medical system that provided the data. The data were obtained retrospectively from administrative records after all clinical decisions had been made, and no information from this study was available to physicians at the time of treatment. While one coauthor served as president of a participating physician group, the data collection process did not influence physician behavior, as all decisions predate the research and were made in the normal course of clinical operations.}

Patient flow is similar across all five EDs. After triage by nursing staff, patients are entered into the EHR tracking board, where they are visible to physicians who are free to self-assign any unclaimed case. Upon assigning one or more patients, a physician conducts initial diagnostics and treatment, which may include ordering laboratory tests, imaging, and medications, and ultimately makes a disposition decision (admit or discharge). Patient assignments are made in a shared central workstation area, which also serves as the hub for chart review and  order entry. During a shift, physicians cycle between the central area and treatment rooms, making rounds on newly assigned and existing patients before returning to the workstation. Travel time between the central area and treatment rooms is minimal (typically less than one minute), which makes it feasible for physicians to switch between clinical and administrative tasks as they see fit.\footnote{In addition to physicians, all EDs in our sample employ physician assistants (PAs). While PAs formally operate under the supervision of physicians and appear as such in the electronic health record system, PAs operate largely independently, primarily managing lower-acuity cases and rarely consulting their supervising physician. Because PA-managed cases follow a distinct workflow and would confound measures of physician workload and assignment behavior, we do not focus on these cases in our analysis.}

\begin{figure}[t!]
\centering
\caption{Physician Case Tracking Board (Generic Sample)} \label{fig:board}
\includegraphics[width=0.85\textwidth]{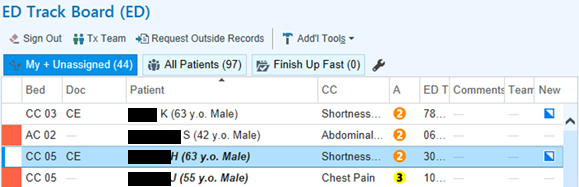}
\end{figure}

When physicians are ready to take on new cases, they can use one of two real-time views of the EHR tracking system; see Figure \ref{fig:board} for a generic screenshot of the interface. The default view ('My + Unassigned') displays the relevant information regarding all cases currently assigned to the physician as well as all unassigned cases waiting in treatment rooms. This includes what bed the patient is assigned to, whether the patient is assigned, the patient's name, age, and gender, the chief complaint, the case acuity, and how long the patient has been in the ED. The secondary view ('All Patients'), which requires clicking on the tab to observe, displays the same details for all cases, including those assigned to other physicians. Our conversations with physicians suggest that physicians typically use the default ('My + Unassigned') rather than the secondary view ('All Patients'). With respect to prioritization, physicians generally give precedence to higher-acuity patients and those that have remained unassigned for longer periods. Finally, there is no stated policy regarding assignment batching. Physicians thus retain full discretion over whether to assign cases individually or to reserve multiple cases at once.


An important feature of the operating environment is physician compensation. Across all physician groups in our sample, physicians are paid a base salary supplemented by a complexity-adjusted piece-rate component. Compensation per case depends on the number of relative value units (RVUs) generated, which are standardized measures used in U.S. healthcare to capture the intensity and resource requirements of clinical services. The specific compensation formula is determined internally by each physician group and is not observable in our data. In particular, the exact weight placed on the RVU-based component is confidential and may vary across groups and over time. While we do not observe physician-level compensation data, at the end of this section we discuss the theoretical implications of this piece-rate scheme for batching behavior, which also motivates our model formulation and experiments.
\subsubsection{Dataset and Sample Construction}

The primary fields used in our analysis include a patient identifier, patient arrival timestamp, arrival-to-treatment-room timestamp, physician assignment timestamp, discharge disposition timestamp, and physician identifier. These variables allow us to identify assignment batching behavior, construct measures of physician workload at the time of assignment, and characterize overall ED congestion. To control for patient-level heterogeneity when capturing the relationship between batching and length of stay (LOS), we additionally observe patient demographic characteristics (legal sex, marital status, race, and age) and clinical details, including chief complaint, Emergency Severity Index (ESI), hospital admission decision, and separate indicators for whether laboratory tests were ordered, whether imaging studies were ordered, and whether medications were administered.

We begin with 1,465,854 cases that were treated by a physician. We remove 9,612 cases missing one or more critical timestamps and 7,278 cases with illogical timestamp sequences (logical sequence: arrival to ED $\rightarrow$ arrival to treatment room $\rightarrow$ disposition decision). Finally, we exclude 6,009 outlier cases with length of stay (defined as the time between arrival and disposition decision) exceeding twelve hours. These reductions result in a patient volume of 1,442,955, ranging between 180,121 and 439,335 cases by hospital. 



\subsection{Model-Free Analysis\label{sec:field:model-free}}
Based on our conversation with our physician coauthor, we define assignment batching as occurring when a physician self-assigns a case within five minutes of their previous assignment as it is generally infeasible for a physician to conduct initial diagnosis and assign a new case in less than five minutes. (We tested alternative cutoffs and they yield similar results.) Note that a case is classified as ``batched'' if either (1) it is assigned within five minutes of the physician's previous case, or (2) the physician's next case is assigned within five minutes after it. That is, if a physician self-assigns case A at 2:00 pm, case B at 2:02 pm and case C at 2:04 pm, all three cases would be classified as batched. 
 
\subsubsection{Assignment Batching Behavior} 
\begin{figure}[t!]
\centering
\caption{Distribution of Physician Batching Propensity by Hospital} \label{fig:physician_batching_histogram}
\includegraphics[width=0.95\textwidth]{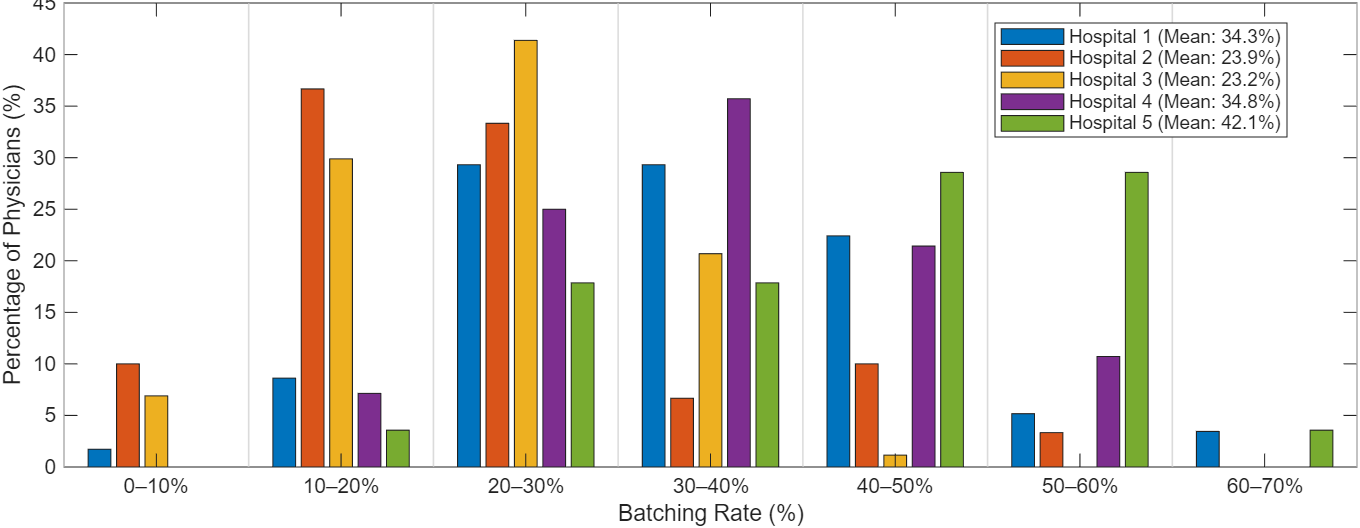}
\end{figure}

We define a batching opportunity as an instance where there are at least two unassigned patients at the time a physician makes an assignment. Across our sample, there are 1,042,665 batching opportunities. We observe substantial variation in batching behavior across physicians and hospitals. Figure \ref{fig:physician_batching_histogram} shows wide dispersion in individual physician batching propensities, ranging from 4.23\% to 65.23\% (median 28.53\%), indicating that some physicians batch very frequently while others rarely do. This heterogeneity persists even within the same hospital, where physicians operate under similar patient populations, staffing levels, and operational protocols, suggesting that differences in individual practice styles play an important role. At the same time, average batching rates differ across hospitals, ranging from 23.2\% to 42.1\%, which suggests that institutional factors such as norms, workflow design, or managerial practices may still play a role. Batching is prevalent across all hospitals in our sample despite differences in compensation schemes, indicating that it is not driven by any particular incentive structure. Taken together, these observations suggest that batching is a widespread but highly variable behavior.
 
Beyond individual and hospital heterogeneity, there is evidence that physicians adjust their batching decisions in response to workload. Table~\ref{tab:batching_by_census_unassigned} presents batching propensity as a function of the physician's own census (the number of cases they are currently managing when choosing whether to batch) and the number of unassigned cases remaining in the ED. Two observations are in order. First, batching propensity decreases in the physician's census. When a physician already has four or more patients, they batch at a rate of approximately 22\% to 30\%, compared with 38\% to 51\% when managing only one patient. This suggests that physicians use batching strategically to regulate their personal workload, i.e., self-assign multiple cases more frequently when their current load is light. Second, batching propensity increases in the number of unassigned cases. When only one unassigned case is available, physicians batch 22\% to 38\% of the time (depending on their own census); when five or more unassigned cases are available, batching rates rise to 25\% to 51\%. This speaks against the incentive motive (i.e., an attempt to increase one's compensation) as being the first order factor driving behavior. If incentives were the primary motive, physicians would want to batch more when the number of unassigned cases is low, to make sure they maintain a high personal workload. Instead, it appears that physicians interpret the presence of multiple unassigned cases as a sign of the ED being congested and respond by batching to ``clear the board'' and help ease congestion. That is, batching may be a well-intentioned but potentially misguided effort to contribute to overall system throughput.\footnote{Appendix~\ref{app:empirics:logit} presents a more formal  analysis of the batching behavior shown in  Table~\ref{tab:batching_by_census_unassigned}. In particular, Table \ref{tab:conditional_logit} shows logit analysis that confirms that, after including physician fixed effects, batching propensity both decreases in the number of currently self-assigned patients and increases in the number of unassigned patients present in the system.}
 
\begin{table}[b]
\centering\footnotesize
\caption{Batching Propensity by Focal Physician Census and Unassigned Cases}
\label{tab:batching_by_census_unassigned}
\begin{tabular}{lccccc}
\hline
Unassigned Cases & Census = 1 & Census = 2 & Census = 3 & Census = 4 & Census = 5+ \\
\hline
1  & 38\% & 27\% & 23\% & 22\% & 22\% \\
2  & 47\% & 32\% & 27\% & 25\% & 25\% \\
3  & 49\% & 35\% & 29\% & 28\% & 28\% \\
4  & 51\% & 37\% & 30\% & 30\% & 29\% \\
5+ & 51\% & 37\% & 32\% & 32\% & 32\% \\
\hline
\end{tabular}
\end{table}

\subsubsection{Batching and Length of Stay}
We also examine the unconditional relationship between batching and patient-level length of stay (LOS). Table~\ref{tab:los_by_batching} reports mean LOS for batched and non-batched cases at each hospital. Across all five hospitals, batched cases exhibit longer stays than single-assignment cases. The magnitude of the difference varies from 5.8 minutes at Hospital 5 to 19.2 minutes at Hospital 3. While these raw differences are suggestive, they do not account for potential confounders: batched cases may differ systematically from non-batched cases in ways that independently affect LOS. Hence, to cleanly identify the relationship between batching and LOS we next control for these variables in our econometric specification.

 
\begin{table}[b!]
\centering\footnotesize
\caption{Mean Length of Stay (Minutes) by Batching Status}
\label{tab:los_by_batching}
\begin{tabular}{ccccc}
\hline
Hospital & All Cases & Batched & Not Batched & Difference \\
\hline
1 & 175.16 & 186.84 & 169.92 & 16.92 \\
2 & 155.31 & 163.68 & 153.26 & 10.42 \\
3 & 197.21 & 212.15 & 192.99 & 19.16 \\
4 & 171.93 & 183.76 & 167.67 & 16.09 \\
5 & 148.31 & 152.56 & 146.73 & 5.82 \\
All & 174.03 & 184.79 & 170.47 & 14.32 \\
\hline
\end{tabular}
\end{table}

\subsection{Impact of Batching on Length of Stay}

\subsubsection{Econometric Specification}
To understand the relationship between batching and LOS we estimate the following linear regression model:
\begin{equation}
\text{LOS}_{ijt} = \beta_1 + \beta_2 \text{Batch\_2}_{ijt} + \beta_3 \text{Batch\_3}_{ijt} + \beta_4 \text{Batch\_4}_{ijt} + \beta_5 \text{Batch\_5+}_{ijt} +\mathbf{X}_{ijt}'\boldsymbol{\gamma} + \alpha_i + \varepsilon_{ijt},
\end{equation} where $\text{LOS}_{ijt}$ is the length of stay (in minutes) for case $j$ treated by physician $i$ at time $t$, $\text{Batch\_b}_{ijt}$ is a binary indicator for whether the case was batched within a batch of $b$ total cases, $\mathbf{X}_{ijt}$ is a vector of control variables, $\alpha_i$ represents physician fixed effects, and $\varepsilon_{ijt}$ is an error term. The coefficients of interest, $\beta_2$-$\beta_5$, measure the within-physician difference in LOS between batched and single-assignment patients, after adjusting for all observed covariates and time-invariant physician characteristics. Note that $\beta_5$ is the effect of any batch that consists of \textit{at least} five cases. 


The controls in $\mathbf{X}_{ijt}$ are as follows. First, we include variables that describe the state of the ED at the time of case assignment: the focal physician's census (number of patients currently assigned to that physician), the focal physician's utilization rate over the preceding eight hours (proportion of time with at least one patient), overall ED busyness (ratio of total active cases to physicians on shift)\footnote{Please see Appendix~\ref{app:empirics:busyness} for the distribution of ED busyness across hospitals. For each of the five hospitals, the modal utilization is between 3 and 5 active cases per physician, and all hospitals see frequent periods of low utilization (0 to 2 cases per physician) and high utilization (more than 6 cases per physician). Overall, there is substantial within- and between-hospital variation, so that the coefficient on ED busyness is well-identified.}, whether the case involved a handoff (transfer from the initial physician to a different concluding physician), and whether the patient was ultimately discharged home rather than admitted to the hospital. Second, temporal controls include categorical variables for year, month, day of week, and hour of arrival to account for seasonality in ED demand and staffing. Third, physician characteristics are absorbed by the fixed effects $\alpha_i$, and we additionally include a continuous measure of each physician's cumulative experience (number of cases previously treated during the study period). Patient demographics include age (in ten-year categories), legal sex, marital status, and race. Fourth, case characteristics are the patient's chief complaint (complaints with fewer than 200 cases are grouped into an ``Other'' category), Emergency Severity Index (ESI) acuity score, number of prior ED visits by the same patient in the past 90 and 365 days, and a categorical variable indicating which combination of diagnostic tests (laboratory, imaging, medications) were ordered during the encounter.
 
\subsubsection{Regression Results\label{sec:field:reg}} Table~\ref{tab:regression_results} presents the estimated coefficients for the main variables of interest across all five hospitals. (Full results including standard errors are reported in Appendix~\ref{app:empirics:regression}). The coefficient on batching is positive and statistically significant across all hospitals and all batch sizes, and the effects exhibit a consistent and striking pattern: the increase in length of stay for batched patients grows monotonically with the number of cases in the batch. At Hospital 1, for instance, a batch of two is associated with a raw increase of 8.86 minutes, rising to 16.14 minutes for batches of three, 26.00 minutes for batches of four, and 35.13 minutes for batches of five or more. Indeed, LOS increases monotonically in the size of the batch across all five hospitals.

\begin{table}[!b]
\centering\footnotesize
\caption{Linear Regression Results: Length of Stay (Minutes)}
\label{tab:regression_results}
\begin{tabular}{lc|ccccc}
\hline
Variable & All Hospitals & Hospital 1 & Hospital 2 & Hospital 3 & Hospital 4 & Hospital 5 \\
\hline
\textit{Batch\_2}                   & 7.46*** & 8.86*** & 5.68*** & 7.61*** & 7.83*** & 5.22*** \\
\textit{Batch\_3}                   & 14.77*** & 16.14*** & 13.20*** & 15.05*** & 13.91*** & 13.23*** \\
\textit{Batch\_4}                   & 22.87*** & 26.00*** & 21.93*** & 20.58*** & 26.10*** & 16.97*** \\
\textit{Batch\_5+}                  & 30.45*** & 35.13*** & 31.78*** & 29.39*** & 36.03*** & 19.11*** \\
\textit{Focal Physician Census}     & 0.88*** & 1.01*** & 0.39*** & 0.68*** & 1.98*** & 1.78*** \\
\textit{ED Busyness}                & 6.65*** & 6.45*** & 5.99*** & 10.65*** & 4.70*** & 4.98*** \\
\textit{Handoff}                    & 16.77*** & 20.24*** & 50.81*** & 10.05*** & $-$3.15*** & 10.38*** \\
\textit{Discharged}                 & 15.95*** & 6.65*** & 14.22*** & 20.28*** & 17.43*** & 14.67*** \\
\hline
Hospital Fixed Effects                       &    \checkmark      &  &  &   &   &   \\
Physician Fixed Effects                        & \checkmark & \checkmark & \checkmark & \checkmark & \checkmark & \checkmark \\
Physician Experience                & \checkmark & \checkmark & \checkmark & \checkmark & \checkmark & \checkmark \\
Physician Utilization               & \checkmark & \checkmark & \checkmark & \checkmark & \checkmark & \checkmark \\
Seasonality                         & \checkmark & \checkmark & \checkmark & \checkmark & \checkmark & \checkmark \\
Patient Demographics                & \checkmark & \checkmark & \checkmark & \checkmark & \checkmark & \checkmark \\
Patient Prior Visits                & \checkmark & \checkmark & \checkmark & \checkmark & \checkmark & \checkmark \\
Chief Complaint                     & \checkmark & \checkmark & \checkmark & \checkmark & \checkmark & \checkmark \\
Acuity Level                        & \checkmark & \checkmark & \checkmark & \checkmark & \checkmark & \checkmark \\
Testing/Medicine Intensity                  & \checkmark & \checkmark & \checkmark & \checkmark & \checkmark & \checkmark \\
\hline
$N$                                 & 1{,}442{,}955 & 342{,}379 & 296{,}458 & 439{,}335 & 184{,}662 & 180{,}121 \\
$R^2$                               & 0.3553 & 0.3582 & 0.3948 & 0.3020 & 0.3517 & 0.3729 \\
Adjusted $R^2$                      & 0.3549 & 0.3575 & 0.3941 & 0.3013 & 0.3507 & 0.3718 \\
\textbf{Weighted Avg \% Effect on LOS} & \textbf{10.09\%} & \textbf{12.21\%} & \textbf{8.78\%} & \textbf{8.75\%} & \textbf{10.36\%} & \textbf{8.94\%} \\
\hline
\end{tabular}
\\[4pt]
\begin{minipage}{0.86\textwidth}
\scriptsize
\textit{Notes.} *** $p < 0.001$. Regression coefficients are reported in minutes. The All EDs specification additionally includes hospital indicator variables. While physician IDs are unique across hospitals, 47 physicians treated cases at more than one hospital in the sample; of these, only 9 treated 500 or more cases at two or more hospitals. These cross-hospital physicians provide the identifying variation for the hospital fixed effects in the pooled specification.
\end{minipage}
\end{table}

\begin{figure}[tbh]
\centering
\caption{Marginal effects of batching on length of stay with 95\% CI.}
\label{fig:batch_margins}
\includegraphics[width=\textwidth]{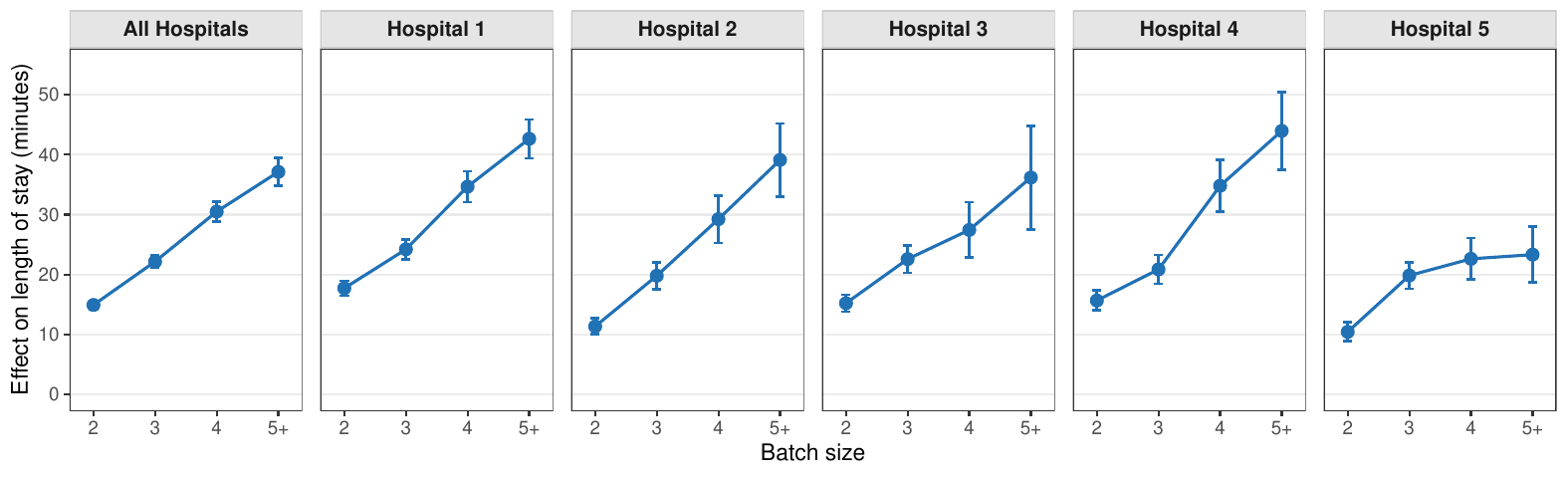}
\end{figure}

These raw coefficients, however, understate the true delay imposed on the average patient in a batch. Because the physicians see patients sequentially, the first patient assigned in any batch faces no additional wait relative to a non-batched case. The delay falls entirely on the remaining patients. To obtain the average effect per affected patient, we compute a modified coefficient that scales each raw estimate by $b/(b-1)$, where $b$ is the batch size, allocating the total delay across only those patients who actually experience it. Figure~\ref{fig:batch_margins} plots these scaled effects. As shown in the figure, a batch of two at Hospital 1 implies an average delay of 17.72 minutes for the patient who waits, a batch of three implies 24.21 minutes per delayed patient, and so on up to 42.64 minutes for batches of five or more. Expressed as a percentage of predicted length of stay for non-batched cases, these effects range from roughly 7\% to 10\% for batches of two, 11\% to 14\% for batches of three, 14\% to 21\% for batches of four, and 16\% to 26\% for batches of five or more. Weighting by the hospital-specific distribution of batched cases across batch-size categories, the average percentage increase in length of stay for batched patients ranges from 8.75\% at Hospital 3 to 12.21\% at Hospital 1. 

The effects of the ED state variables are also similar across hospitals. Higher physician census is associated with longer length of stay, with coefficients ranging from 0.39 to 1.98 minutes per additional patient in the physician's workload. ED busyness is positively associated with length of stay in all hospitals, with coefficients ranging from 4.70 to 10.65 minutes per unit increase in the busyness ratio, consistent with congestion delays. Handoffs are associated with longer length of stay in four of the five hospitals; Hospital 4 is an exception, where the coefficient is negative, possibly reflecting idiosyncratic handoff practices or shift structures at that site. Discharged patients have longer stays than admitted patients in all five hospitals, with coefficients ranging from 6.65 to 20.28 minutes. Overall, the model explains a substantial portion of the variance in length of stay, with $R^2$ values ranging from 0.30 to 0.39 across hospitals.

\subsubsection{Alternative Explanations}
The main threat to interpreting the batching coefficient as the effect of batching itself is that batching may be correlated with other factors that also increase length of stay. ED busyness is the most obvious factor: a congested system has more unassigned cases and hence more opportunities to batch, while simultaneously driving up length of stay. Our regression therefore controls for ED busyness. Similar reasoning applies to physician census (number of patients currently treated): physicians who systematically prefer heavier caseloads may both batch more and process cases more slowly. Physician fixed effects address the possibility that physicians who batch more are also systematically slower processors of cases. Finally, the combination of diagnostic tests, imaging, and medications ordered during an encounter captures case complexity in a way that may be correlated with both batching and longer stays. The fact that the batching coefficient remains positive, statistically significant, and similar in magnitude across all five hospitals after including these controls gives us confidence that the effect is real. The monotonic increase in length of stay with batch size ($\beta_5 > \beta_4 >\beta_3 >\beta_2 >\beta_1$) further reinforces this, since each additional case in a batch must wait for more cases to be treated before its own treatment can begin.

A remaining potential concern is selection on unobservables, in particular that physicians may batch cases based on information that is available to them at the moment of assignment but is unobservable in the data record. If physicians systematically batch cases that are more complex on unobservable dimensions, this would inflate our estimates. Two reasons speak against this concern. First, assignment decisions are made at the central workstation terminal, typically before the physician has examined the patient. The information available to the physician at that moment (chief complaint, acuity score, time in the ED, and basic demographics) is captured in our controls, leaving little room for selection on unobserved complexity. Second, even if such information were available, prior evidence on case selection in EDs suggests that selection would run the other way. \citet{KCStaatsKouchakiGino2020} show that physicians tend to select easier cases under congestion, the same conditions in which batching concentrates in our data. To the extent that unobserved selection on case complexity is occurring, it would therefore push the batching coefficient toward zero rather than inflate it, so our estimates are, if anything, on the conservative side.
 
\subsection{Motivation for Model and Experiments}
The regression results in \textsection\ref{sec:field:reg} show that batching is associated with significantly longer stays for the batched patients, with effect sizes of 10 to 44 minutes per batched case depending on hospital and batch size. It is worth emphasizing that our estimates are patient-level (as opposed to system-level). That is, we identify how much longer a batched patient stays in the ED relative to a comparable single-assignment patient seen by the same physician. Indeed, with the available data we are unable to estimate the effect of batching on overall ED throughput. The estimates do, however, strongly suggest that batching increases the \textit{variance} of length of stay across patients, since patients at the back of a batch wait longer than those ahead of them or those that were assigned one at a time. Given that LOS variability is a key metric for most hospitals \citep{Welch2011EDMetrics,Mason2012FourHour}, the variance increase alone would mean a substantive operational cost.

To design an effective remedy for batching, we first need to understand \textit{why} physicians batch. We believe that there are two plausible explanations. First, all five hospitals in our sample use some form of piece-rate compensation, where physicians' earnings depend at least partially on the number of patients treated. Under such incentive schemes, physicians may use assignment batching as a form of ``hoarding'' patients to maximize personal utilization. By reserving multiple cases at once, a physician reduces the risk of finishing all assigned work and subsequently sitting idle while waiting for the next case. That is, batching may be an individually optimal response to financial incentives. Second, the fact that batching increases with the number of unassigned cases suggests that physicians may use batching as a strategy to reduce ED congestion. Physicians may interpret a long queue of unassigned patients as a signal of system stress and respond by batching to ``clear the board'' more quickly. By taking on more work themselves, physicians may believe they are freeing up capacity for their colleagues and reducing overall wait times. Rather than being selfish, batching may therefore have an altruistic (but potentially misguided) motive to help system performance.  

To separate the two mechanisms, we next turn to a stylized model that gives us a prediction for how a rational, profit-maximizing physician should behave under a given compensation structure. We then conduct controlled experiments in which we vary the incentive system across treatments. Under individual throughput incentives, the payoff-maximizing strategy should be to batch; under group incentives that reward overall system performance, the optimal strategy should be to assign one patient at a time. If participants follow the incentives (batch under individual, not batch under group incentives), then redesigning physician compensation is likely to be an effective remedy. If, in contrast, participants continue to batch even under group incentives, then batching would appear to be a  behavioral tendency to prioritize personal workload and busyness over collective efficiency. In the latter case, compensation alone is unlikely to shift behavior, and changes to the assignment interface and to the information available to physicians would be needed. We next develop a stylized model of self-assignment decisions in an emergency department with stochastic arrivals, stochastic treatment times, finite capacity and multiple physicians. 

\section{Model}\label{sec:Model}
Many mechanisms could plausibly explain why physicians batch: responding to incentives (RVU-based compensation), strategic behavior towards colleagues, team norms, a preference for keeping busy rather than idle or a belief that batching helps increase patient throughput. We focus on the individual-productivity and workload-management mechanism because it is one that most closely aligns with (a) what is suggested by our field data and (b) anecdotal evidence from interactions with physicians across the EDs in our dataset. Accordingly, both our model (\textsection 4) and our experiments (\textsection 5) hold partner behavior fixed and common knowledge, reducing the focal physician's problem to a single-agent dynamic optimization problem under uncertainty.

\subsection{Model Description}

\subsubsection{Resolution Process}

Patients arrive at an emergency department with $M$ treatment rooms according to a Poisson process with rate $\lambda$. Each arrival consists of a random number of patients $K$, supported on $\{1,2,\dots,B\}$, with probabilities $\mathbb{P}(K = k) = p_k$ for $k = 1,\dots,B$. Let $P = (p_1,\dots,p_B)$ denote the arrival-size distribution. If an arriving group exceeds the number of available treatment rooms, rooms are filled to capacity and any remaining patients are blocked and diverted elsewhere.\footnote{Patient blocking is a common simplification in queueing models of healthcare operations \citep{deBruin2007cardiac, deBruin2010erlang, Asaduzzaman2010loss}. It keeps the state space finite, which makes the model analytically tractable and makes the experimental setup (in \textsection 5) simpler for participants. The disadvantage  is that it abstracts away from a common waiting area. Note, however, that, if anything, arrival blocking limits the externalities of suboptimal assignment policies as it caps the maximum queue length.}

A total of $N \geq 2$ physicians, indexed by $i \in \{1,\dots,N\}$, treat patients in the $M$ shared treatment rooms. Newly arrived patients enter one of the unoccupied treatment rooms as unassigned and remain so until claimed by a physician. At each decision epoch (which will be formally defined later) a physician can claim one or multiple unassigned patients. Once assigned, a patient cannot be reassigned and exits the system only after completing treatment with that physician. Upon completing all assigned cases, a physician returns to a central terminal. Since physicians are homogeneous with respect to their service time distributions and delaying service initiation while unassigned patients are present is wasteful, idling cannot be system optimal. Accordingly, if unassigned patients are present upon return, the physician must immediately self-assign at least one case, but may claim any or all unassigned cases. If no unassigned patients are present, the physician remains idle at the terminal until the next arrival, at which point they must immediately self-assign at least one case. After self-assigning one or more cases, a physician administers treatment to each assigned patient. Treatment times are independent and exponentially distributed with mean $1/\mu$. Patients exit their treatment rooms immediately upon completion of service, and the physician returns to the central terminal once all assigned patients are treated.

\subsubsection{States and Strategies}

Let $A_i(t)$ denote the number of cases assigned to physician $i$ at time $t$, where $A_i(t) = 0$ indicates that physician $i$ is idle at the terminal. The state of the system at time $t$ is given by $(U(t), \mathbf{A}(t))$, where $U(t)$ is the number of unassigned patients currently occupying treatment rooms and $\mathbf{A}(t) = (A_1(t),\dots,A_N(t))$ is the vector of physician caseloads. All physicians follow non-idling strategies (as explained in the previous paragraph), so decision epochs occur only when a physician returns to the terminal and more than one unassigned patient is present, or when a physician is idle at the terminal and an arriving group contains more than one patient. 

Each physician follows a stationary strategy that depends only on the current system state. Let $d_i(U,\mathbf{A})$ denote physician $i$'s decision rule specifying how many unassigned cases to self-assign given the system state. The strategy of physician $i$, denoted by $D_i$, is the collection of decision rules $d_i(U,\mathbf{A}) \leq U$ for all states with $A_i = 0$ and $U \geq 2$. Let $\mathbf{D} = (D_1,\dots,D_N)$ denote the strategy profile of all physicians. Given the finite state space and the exponential arrival and treatment assumptions, performance measures including individual physician throughput, system throughput, and expected patient sojourn time can be evaluated for any given strategy profile $\mathbf{D}$ and parameter set $\mathbf{\Theta} = \{M,N,\lambda,P,\mu\}$ by characterizing the system as a CTMC and computing the steady-state probabilities over all states $(U,\mathbf{A})$.

\subsection{Focal Physician Decision Problem}
We examine the patient assignment problem from the perspective of an individual physician whom we refer to as the \textit{player} ($i=1$). We refer to the remaining $N-1$ physicians as the \textit{partners}. To focus on individual self-assignment behavior, partner behavior is treated as exogenous: partner strategies, denoted $\mathbf{D}_{-1}$, are fixed, known to the player, and do not depend on the player's own strategy $D_1$.

\subsubsection{Optimal Strategy Under Personal Throughput Incentives}
In our empirical setting (in \textsection 3), physicians across all five hospitals are compensated at least partially through a piece-rate component, which may incentivize them to maximize personal throughput. The exact form of the optimal individual strategy under such incentives depends on the arrival process, the system capacity, and the behavior of partners. For example, under an arrival process that mixes  singletons with occasional bursts, it can be individually optimal to assign one patient at a time during quiet periods and to batch only after a burst arrives and multiple unassigned patients are simultaneously available. The optimal strategy can also respond to partner behavior. For example, if partners batch more aggressively, the focal physician's best response can also be more aggressive. Rather than characterize the full set of optimal individual strategies across all parameter configurations, we focus on the configuration we use in the experiment (in \textsection 5). The analysis for this case is clean and the prediction is testable. In particular, the experiment features groups of three patients arriving at rate $\lambda$, a single partner (i.e., $N=2$) who always assigns exactly one patient, a capacity of $M=4$ treatment rooms, and a maximum batch size of two. Under these conditions, the following result holds.

\begin{proposition}\label{prop:personal} Under personal throughput incentives, with groups of three patients arriving at rate $\lambda$, $N=2$, a partner who always assigns one patient, $M=4$ treatment rooms, exponential service times with rate $\mu$, and a maximum batch size of two, the batching policy is the unique optimal strategy for the focal physician for all $\lambda, \mu > 0$. \end{proposition}

The proof, provided in Appendix~\ref{app:proofs}, solves the Poisson equations for the average-reward CTMC under the batching policy and verifies that the resulting relative value function certifies optimality at every decision state. The relative value function $h(s)$ measures the expected cumulative personal throughput advantage of starting in state $s$ relative to the empty and idle state $(0,0,0)$, with the long-run average rate subtracted out. Decision epochs arise in states $(0,0,0)$ and $(2,1,1)$, and at both states the choice reduces to the same comparison. Self-assigning two patients transitions the system to state $(0,2,1)$, while self-assigning one patient allows the partner to claim the second, transitioning to $(1,1,1)$. The appendix shows that $h(0,2,1) > h(1,1,1)$ for all $\lambda, \mu > 0$, confirming that batching uniquely maximizes the focal physician's personal throughput at every decision epoch under any arrival and service rate.

The intuition behind Proposition 1 is straightforward. A case already assigned to the focal physician guarantees maximum individual throughput (in the short term), whereas an unassigned case is at risk of being claimed by the partner at their next decision epoch. State $(0,2,1)$, in which the focal physician has self-assigned two cases, is therefore strictly more valuable than state $(1,1,1)$, in which the focal physician has self-assigned only one case and a second remains unclaimed, for any combination of arrival and service rates. This is because under exponential service times the focal physician and the partner are equally likely to be the next to complete a case. In state $(1,1,1)$ the partner therefore claims the unassigned case with probability one half before the focal physician has a chance to do so. Self-assigning the second case immediately and moving to state $(0,2,1)$ eliminates this possibility.

\subsubsection{Optimal Strategy Under Team Performance Incentives}
While personal throughput incentives make batching individually optimal for the parameters used in our experiments (Proposition 1), under team incentives we are able to prove a more general result, with an arbitrary number of treatment rooms and physicians. Specifically, we focus on two system-level measures: system throughput and expected patient sojourn time. Let $\mathbb{E}[W \mid \mathbf{\Theta}, \mathbf{D}]$ denote the expected sojourn time and $\Lambda(\mathbf{\Theta}, \mathbf{D})$ the long-run system throughput under parameters $\mathbf{\Theta}$ and strategy profile $\mathbf{D}$.

\begin{proposition}\label{prop:system} Under team performance incentives, where each physician's payoff is increasing in system throughput or decreasing in expected patient sojourn time, the strategy that maximizes each individual physician's payoff is to assign exactly one patient at each decision epoch, since long-run system throughput $\Lambda(\mathbf{\Theta}, \mathbf{D})$ is maximized and expected sojourn time $\mathbb{E}[W \mid \mathbf{\Theta}, \mathbf{D}]$ is minimized when every physician assigns exactly one case at each decision epoch; that is, when $d_i(U,\mathbf{A}) = 1$ for all $i, U, \mathbf{A}$.
\end{proposition}

The proof, provided in Appendix~\ref{app:proofs}, uses a sample-path coupling argument that fixes a realization of arrivals and a sequence of potential service requirements revealed upon service initiation, and shows that any policy under which a physician assigns more than one patient (weakly) delays service initiation for at least one patient relative to the always-assign-one policy. These delayed initiations propagate to (weakly) later completion times, and a stochastically larger number of patients in the system at all subsequent times, which in turn leads to lower throughput through increased blocking, and higher sojourn times through increased congestion. It follows that since assigning one patient at a time maximizes system throughput and minimizes sojourn time for all physicians simultaneously, no individual physician can improve their own payoff by batching, and the individually optimal strategy coincides with the system-optimal strategy.\footnote{We remark that Proposition~\ref{prop:system} holds for general inter-arrival and service time distributions. The sample-path coupling argument does not rely on the Poisson arrival or exponential service assumptions of the CTMC model; it requires only that service times be independent across patients and revealed at service initiation.}

\subsection{Motivation for Experiments}
To test the predictions of Propositions~\ref{prop:personal} and~\ref{prop:system}, we next turn to controlled  experiments. The model was designed with these experiments in mind, with simplifications chosen so that the decision situation can be recreated in a short session with busy medical professionals. In particular, the finite queue assumption keeps the state space small and finite (only nine reachable states in the CTMC; see Appendix~\ref{app:proofs}) and avoids the more complex queueing dynamics that would arise if arriving patients in excess of capacity could wait outside the treatment area. Similarly, treating partner behavior as exogenous removes the strategic interaction between physicians that would otherwise complicate the focal physician's optimization problem. 

While a more general model would be interesting to analyze, characterizing optimal behavior under richer assumptions is not the focus of this paper. Rather, we are interested in whether medical professionals' behavior aligns with the model predictions. The answer to this question is central to the design of interventions to reduce batching. If batching is a rational response to piece-rate compensation, an appropriate intervention is to redesign physician contracts. If batching continues to persist even after the piece-rate incentive has been removed, then redesigning physician compensation alone is likely not going to be sufficient and changes to the assignment interface in the electronic health record system (i.e., how information is presented to physicians when making patient assignment decisions) would be needed. The field data alone (\textsection 3) cannot answer this question, since we do not observe compensation structure, and since the interface (EHR patient tracker) is the same across all EDs. We next introduce controlled online experiments in which we test Propositions~\ref{prop:personal} and~\ref{prop:system} using exogenous variation in the compensation structure.

\section{Experiments}\label{sec:Experiment}
To test whether medical professionals behave in line with the theory developed in \textsection 4, we ran two pre-registered experiments: one with a targeted sample of medical professionals recruited on Prolific, and a second one with verified ED physicians. Participants played the role of the focal physician in the ED system described in the theoretical model in \textsection 4.  Please see Appendix~\ref{app:experiment} for pre-registration details, exclusions, participant demographics, experiment instructions, comprehension checks and other details.

\subsection{Experiment 1: Medical Professionals Sample (Prolific)}
We recruited $229$ Prolific participants currently employed in the medical/healthcare sector (according to Prolific's own determination), residing in the US or Canada, with a target of $N=50$ per treatment.\footnote{At the start of the experiment, participants additionally self-identified their healthcare role from the following categories: physician/doctor, nurse, paramedic/EMT, pharmacist, therapist (physical, occupational, etc.), medical technician/lab tech, healthcare administrator, or other clinical role. We excluded psychologists and social workers from being eligible to sign up because their work is typically appointment-based, i.e., does not involve dynamic patient assignment decisions. Therapists (physical, occupational, speech-language, and respiratory) were retained because they often work in hospital settings where allocation of patients from a shared consult or referral pool is a routine part of practice. An additional $11$ participants were screened out at the beginning of the experiment for lacking healthcare experience (despite being classified as healthcare workers by Prolific).} Of these, $26$ did not pass our pre-registered comprehension and quality checks and were excluded ($21$ based on comprehension test, $3$ for excessive completion time, and $2$ for reporting technical difficulties; see Appendix~\ref{app:experiment} for details), yielding a final analysis sample of $N=203$ participants (between $47$ and $53$ per treatment).\footnote{The mean age of the analysis sample is 39.2 years ($\text{SD} = 11.6$; range 20--72), with 71.4\% being female. The most commonly reported healthcare roles are nurse (57), other clinical staff (38), therapist (19), technician (17), administrator (15), and physician (14); the remaining participants hold other clinical roles such as paramedic or pharmacist. All participants passed the Prolific employment screener confirming current employment in the medical or healthcare sector. See Appendix \ref{app:demographics} for details.}

\subsubsection{Experiment Design}

The focal choice in the experiment is between assigning one vs. two patients (batching). At each decision epoch, participants see a waiting room displaying the currently unassigned patients (panel~(a) of Figure~\ref{fig:exp_interface}). Patients arrive in groups of three at exponentially distributed intervals ($\lambda=1/30$), and each patient requires an exponentially distributed treatment time ($\mu=1/15$). The system has two servers ($N=2$) and a total patient capacity of four ($M=4$). This means that when two or more patients are waiting to be treated, arriving patients in excess of available capacity are turned away. After the assignment decision, the participant treats assigned patients sequentially, seeing a progress bar for each patient fill according to the realization of the random treatment time, with five time units corresponding to one second (panel~(b) of Figure~\ref{fig:exp_interface}).


Even though there are two physicians on duty, our experimental task is an individual decision-making problem. In particular, the computerized partner is programmed to self-assign exactly one patient (and this is communicated to the participant). Because the partner's behavior is fixed and non-strategic, the task reduces to a single-agent optimization problem. This allows us to cleanly identify whether participants choose the most productive strategy under each incentive system, or whether they exhibit a bias in making assignment decisions. After self-assigning a new patient the partner begins treating the patient (with the same exponential service rate of $\mu=1/15$ as the focal player). Partner strategy, characteristics (service rate) and behavior are common knowledge and are shared with the participant. Each shift lasts 600 time units. Terminal conditions that ensure stationarity of the optimal policy throughout the horizon are derived in Step 2 of Proof 1 in Appendix~\ref{app:proofs}; participants received a plain-language explanation of these conditions as part of the instructions. 

\subsubsection{Experimental Procedure}
\begin{figure}[p]
\centering
\caption{Screenshots of the experimental interface.}
\label{fig:exp_interface}
\begin{subfigure}[t]{0.67\textwidth}
\centering
\caption{Patient list and assignment decision (Practice round, \textbf{\textit{GT}} treatment).}
\label{fig:exp_patient_list}
\includegraphics[width=\textwidth]{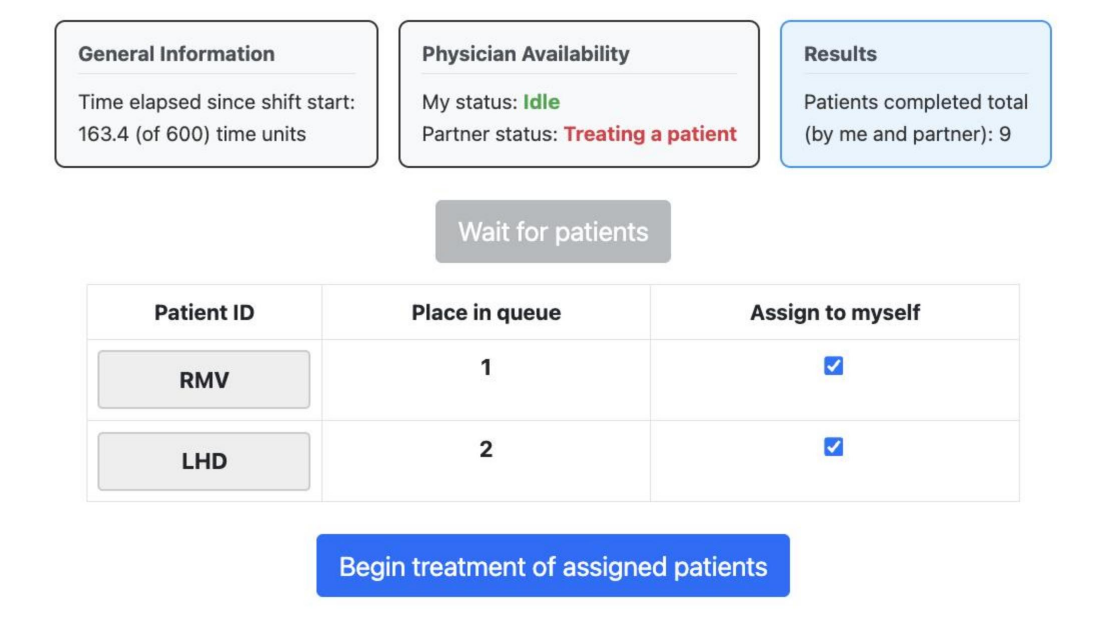}
\end{subfigure}\\[16pt]
\begin{subfigure}[t]{0.64\textwidth}
\centering
\caption{Treatment room (Practice round, all treatments).}
\label{fig:exp_treatment_room}
\includegraphics[width=\textwidth]{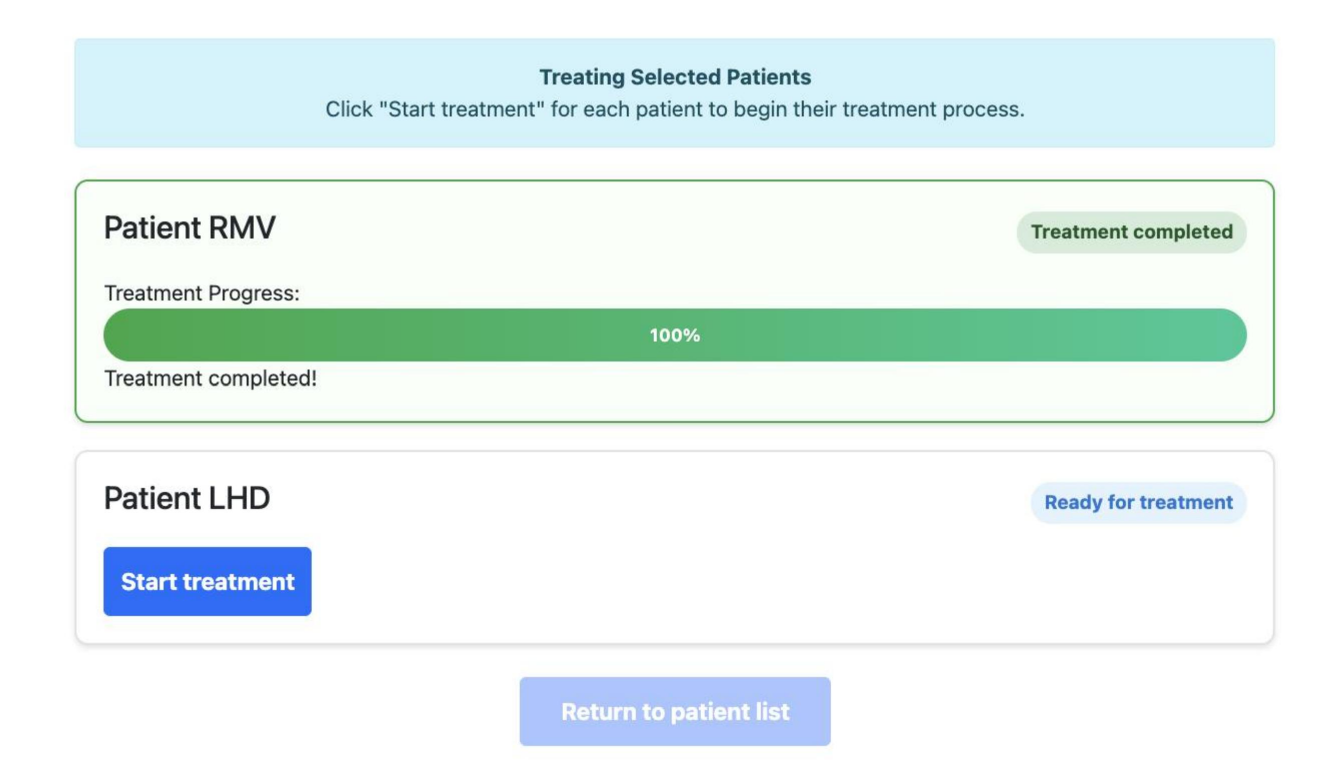}
\end{subfigure}\\[16pt]
\begin{subfigure}[t]{0.73\textwidth}
\centering
\caption{Strategy commitment screen (Main round, \textbf{\textit{GT}} treatment).}
\label{fig:exp_strategy_group}
\includegraphics[width=\textwidth]{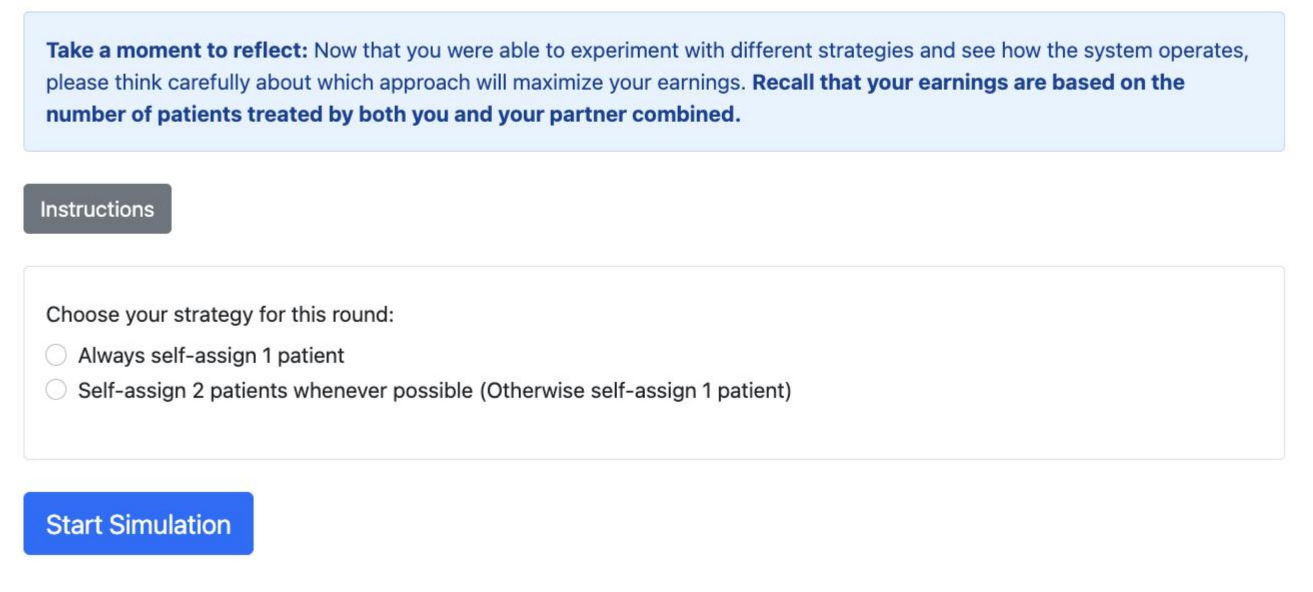}
\end{subfigure}
\smallskip
\begin{minipage}{0.92\textwidth}
\scriptsize
\textit{Notes.} Panels (a) and (b) show the practice-round screens (decision and patient treatment screens). Panel (c) shows the Round 2 strategy-commitment screen. All screen shots are from the \textbf{\textit{GT}} (group throughput) treatment.
\end{minipage}
\end{figure}

After completing the instructions and comprehension screening, participants complete two rounds of 600 time units each. In the first (practice) round, participants repeatedly make self-assignment decisions after observing the current system state. Panel~(a) of Figure~\ref{fig:exp_interface} shows one such state, with 163.4 time units having elapsed since the start of the round, the participant at the terminal facing a decision, and the partner treating a patient. A decision is made either when the participant returns from the treatment room after completing all assigned patients, or when a new group of patients arrives while the participant is idle at the terminal. The purpose of the practice round is to let participants experience the system dynamics before committing to a strategy in Round~2. Each participant makes between 10 and 20 decisions over the course of the round, depending on their strategy and the realization of arrivals. At the end of the round the participant is shown their total payoff. 

The second (incentivized) round consists of a single decision that is the main object of our analysis. Participants commit at the start of the round to one of two fixed strategies, Always Self-Assign 1 or Always Self-Assign 2 (whenever possible), as shown in panel~(c) of Figure~\ref{fig:exp_interface}. The selected strategy is then executed automatically for the rest of the round. This single-decision format identifies each participant's preferred policy and separates it from any learning, experimentation, or strategy changes that might occur during live play. The format is a variant of the \emph{strategy method} \citep{brandts2011strategy}, which has been shown to elicit behavior similar to direct play across a wide range of experimental settings while substantially reducing within-participant noise. The strategy method is especially well suited to sequential decision problems like ours, where live play can make it difficult to identify the strategy used by the participant \citep{kagan2025sequential}.

\subsubsection{Treatments}
We administered four between-subject treatments, with subjects assigned to one of the four treatments at random upon signing up to the experiment. The main difference between the treatments was how participants were compensated. In particular:

\begin{itemize}[leftmargin=20pt]
\item\textbf{\textit{IT} (Individual Throughput):} Participants are compensated based on individual throughput (the number of patients they personally treat). This creates an incentive to batch (self-assign 2 patients).

\item\textbf{\textit{GT} (Group Throughput):} Participants are compensated based on joint throughput (the total number of patients treated by both the participant and their partner). This creates an incentive to not batch.

\item\textbf{\textit{GT-Nudge} (Group Throughput + Nudge):} Identical to \textbf{\textit{GT}}, but participants see an additional sentence prior to making decisions that reads: \textit{``If you choose Strategy 2 (Self-assign 2 patients whenever possible), the following can occur: While you are treating your two patients, your partner becomes idle and is not treating any patients.''}

\item\textbf{\textit{GT-ST} (Group Throughput based on Sojourn-Time):} Participants are compensated based on total sojourn time for all patients entering the system. This treatment is parametrized to create incentives that are identical to the \textbf{\textit{GT}} treatment since throughput can be mapped one-to-one to total sojourn time.
\end{itemize}

\subsubsection{Incentives}
To ensure that any observed treatment differences are driven by differences in how participants respond to the incentive structure (rather than differences in the monetary stakes or costs of deviating from the optimal policy), we carefully calibrate parameters in each treatment. Doing this in a finite horizon setting that is amenable to a relatively short human-subject experiment is non-trivial, and we describe the detailed procedure in Appendix~\ref{app:experiment:incentives}. We use the transient analysis of the underlying continuous-time Markov chain to select parameters that make both expected earnings and incentive strength (the percentage increase in expected bonus from following the optimal strategy relative to the non-optimal policy) approximately equal, at about 15\% across all four treatments. Because there are only 10 to 20 decisions per round, there is substantial sample path variance; we therefore additionally verify that the calibration holds (i.e., that incentive strength continues to be close to 15\%) for the specific sample paths used in the experiment. 

In all treatments, participants received a base fee of \$2.00 plus a performance-based bonus from the incentivized round (a bonus of \$3.96 on average). Mean completion time was 15.2 minutes, resulting in an average hourly rate of \$23.59. Please see Appendix~\ref{app:experiment:incentives} for the calibration of incentive parameters, selection of sample paths, terminal conditions and other details. 

\subsubsection{Hypotheses}
Our first three hypotheses focus on whether decision-makers correctly incorporate economic incentives into their decisions. Under individual throughput incentives (\textbf{\textit{IT}}), batching is the payoff-maximizing strategy (Proposition 1) so standard incentive theory predicts high batching rates. Under group throughput incentives (\textbf{\textit{GT}}), not batching maximizes team output and hence the participant's bonus (Proposition 2). Because decisions may be noisy rather than perfectly aligned with theory, our hypotheses are based on directional effects. In particular, we test whether batching rate under each incentive system is significantly different from 50\% (Hypotheses 1a and 1b), as well as whether batching rates in the two systems (\textbf{\textit{IT}} and \textbf{\textit{GT}}) are significantly different from each other (Hypothesis 2).

\smallskip

\renewcommand{\thehypothesis}{1a}
\begin{hypothesis}\label{hyp:incentives_batch}
Under individual throughput incentives (\textbf{\textit{IT}}), participants will batch more than 50\% of the time.
\end{hypothesis}

\renewcommand{\thehypothesis}{1b}
\begin{hypothesis}\label{hyp:incentives_no_batch}
Under group throughput incentives (\textbf{\textit{GT}}), participants will batch less than 50\% of the time.
\end{hypothesis}

\renewcommand{\thehypothesis}{2}
\begin{hypothesis}\label{hyp:relative}
Batching rates will be lower under group throughput incentives (\textbf{\textit{GT}}) than under individual throughput incentives (\textbf{\textit{IT}}).
\end{hypothesis}
\smallskip

\noindent While we expect batching to be an intuitive response under individual incentives (\textbf{\textit{IT}}), not batching and responding optimally to group incentives (\textbf{\textit{GT}}) requires that subjects understand the dynamics of the service system (i.e., that holding fewer patients frees cases for the partner, whose output also counts toward the bonus). Prior experimental work suggests this logic may not be straightforward to decision-makers. Subjects often deviate from optimal policies in sequential decision problems \citep{seale1997sequential, kagan2025sequential} and tend to under-weight the system-level effects of their own decisions in healthcare queueing settings \citep{kim2020admission, kremer2023mismanaging}. Further arguments against Hypotheses 1a, 1b, and 2 can be found in the psychology literature that shows that humans often prefer action over inaction to avoid idle time and may treat busyness as inherently valuable, even when inaction is objectively optimal \citep{bareli2007action, hsee2010idleness, bellezza2017conspicuous}; we review this literature in detail in \textsection\ref{sec:Literature}. Thus, decision-makers may default to batching regardless of the incentive structure.

We also test two targeted interventions that may help decision-makers recognize the (negative) effects of batching on system performance. The first  intervention is an information nudge (\textbf{\textit{GT-Nudge}}), where participants face the same group throughput incentives as in \textbf{\textit{GT}} but are additionally told that batching can leave their partner idle while the participant has two cases assigned. Pointing to such system-level inefficiencies should reduce batching. On the other hand, if the underlying driver is not an informational gap but idleness aversion \citep{hsee2010idleness}, then the nudge may have limited effect; participants may fully understand that batching may idle the partner and still prefer it because not batching means accepting personal idle time. The second intervention (\textbf{\textit{GT-ST}}) changes the performance metric from the number of patients treated to total patient sojourn time. While both the number of patients treated (\textbf{\textit{GT}}) and patient sojourn times (\textbf{\textit{GT-ST}}) are calibrated such that they lead to exactly the same incentives (see \ref{app:experiment:incentives}), we expect the latter to make the cost of congestion more salient, and therefore to reduce batching.

\smallskip

\renewcommand{\thehypothesis}{3a}
\begin{hypothesis}\label{hyp:nudge}
Adding a nudge that highlights the consequences of batching (\textbf{\textit{GT-Nudge}}) will reduce batching rates relative to the same incentive structure without the nudge (\textbf{\textit{GT}}).
\end{hypothesis}

\renewcommand{\thehypothesis}{3b}
\begin{hypothesis}\label{hyp:sojourn}
Batching rates will be lower when group incentives are implemented via patient sojourn times (\textbf{\textit{GT-ST}}) than via group throughput (\textbf{\textit{GT}}).
\end{hypothesis}



\subsubsection{Hypothesis tests}\label{subsec:Prolific:Results}
Per our pre-registration, we use one-sided tests of proportion to test Hypotheses 1a, 1b, 2, 3a, and 3b. Recall that the main outcome variable is the strategy chosen in Round 2 (the incentivized round): always self-assigning one patient (``Never Batch'') or always self-assigning two patients when possible (``Always Batch''). Table~\ref{tab:prolific_results} reports the batching rate (the fraction of participants who chose the ``Always Batch'' strategy in the incentivized round) by treatment. Under \textbf{\textit{IT}}, where batching is the payoff-maximizing strategy, 94.3\% of participants chose to batch, significantly above the 50\% benchmark (test of proportions, $p < 0.001$), supporting Hypothesis 1a. Under \textbf{\textit{GT}}, where \emph{not} batching maximizes the participant's bonus, 94.1\% of participants nonetheless chose to batch (test of proportions against 50\%, $p > 0.999$), opposite to the predicted direction; Hypothesis 1b is not supported. The difference between \textbf{\textit{GT}} and \textbf{\textit{IT}} is not statistically significant (test of proportions, $p = 0.481$); Hypothesis 2 is not supported. Switching from individual to group throughput incentives has essentially no effect on behavior: participants batch at the same rate regardless of whether their bonus depends on their own output or the team's.

\begin{result}\label{res:incentives}
Participants respond to individual throughput incentives by batching at high rates (Hypothesis 1a supported). However, group throughput incentives do not reduce batching (Hypotheses 1b and 2 not supported): batching rates under \textbf{\textit{GT}} are nearly identical to those under \textbf{\textit{IT}}.
\end{result}

Adding the nudge message in \textbf{\textit{GT-Nudge}} reduces the batching rate to 76.9\%, compared with 94.1\% under the otherwise identical \textbf{\textit{GT}} treatment. This 17.2 percentage-point reduction is statistically significant (test of proportions, $p = 0.014$). Under \textbf{\textit{GT-ST}}, where earnings depend on patient sojourn times rather than group throughput, the batching rate is 91.5\%, not significantly different from \textbf{\textit{GT}} (test of proportions, $p = 0.455$). Thus, reframing the performance metric from group throughput to patient sojourn times does not reduce batching. Even with the nudge, which is the most effective intervention, more than three-quarters of participants still choose to batch.

\begin{result}\label{res:debiasing}
Making it salient that batching may lead to partner idleness significantly reduces batching relative to group incentives alone (Hypothesis 3a supported). Reframing incentives in terms of patient sojourn times does not reduce batching (Hypothesis 3b not supported).
\end{result}

\begin{table}[t]
\centering\footnotesize
\caption{Batching Behavior by Treatment (Prolific Sample)}
\label{tab:prolific_results}
\renewcommand{\arraystretch}{1.15}
\begin{tabular}{lcccc}
\hline
Treatment & $N$ & Never Batch & Always Batch & Batch Rate (\%) \\
\hline
\textbf{\textit{IT}}     & 53 & 3  & 50 & 94.3 \\
\textbf{\textit{GT}}         & 51 & 3  & 48 & 94.1 \\
\textbf{\textit{GT-Nudge}}   & 52 & 12 & 40 & 76.9 \\
\textbf{\textit{GT-ST}}  & 47 & 4  & 43 & 91.5 \\
\hline
Total & 203 & 22 & 181 & 89.2 \\
\hline
\end{tabular}
\end{table}

\subsubsection{Additional Results}\label{subsec:Prolific:Add:Results}
The persistence of batching under group incentives is difficult to reconcile with standard incentive theory. The decision is binary, so participants are not failing to correctly calibrate their responses. Comprehension is verified through a comprehension quiz and reinforced in the practice round, so participants are not failing because they misunderstand the instructions. Quiz errors are uncorrelated with optimal play ($\rho = 0.09$, $p = 0.19$), and restricting the analysis to the 138 participants who answered all comprehension questions correctly on the first attempt leaves the main results unchanged: $91.7\%$ of these participants still chose to batch under \textbf{\textit{GT}}, and the nudge effect in \textbf{\textit{GT-Nudge}} remains significant ($p = 0.023$). Finally, partner behavior is fixed and common knowledge, so there is no strategic uncertainty about what others might do. 

Results 1 and 2 are also robust to regression analysis that includes a variety of demographic controls (pre-registered as part of our follow-up analysis). Table~\ref{tab:logit_prolific} reports logit regressions of the optimal strategy choice on treatment dummies, demographics, and professional experience variables. Across all specifications, the treatment effects are stable: \textbf{\textit{GT-Nudge}} is the only treatment that significantly increases optimal play relative to \textbf{\textit{GT}} ($p < 0.05$ in all specifications), while \textbf{\textit{GT-ST}} is not significantly different from \textbf{\textit{GT}}. Demographic variables (age, gender, education) are not significant predictors of optimal play in the baseline specifications. Among the remaining variables that we collected, clinical experience predicts optimality ($p < 0.05$ in the full sample); however, even for the more experienced participants (based on median split), batching rate is approximately 85\% when pooling \textbf{\textit{GT}}, \textbf{\textit{GT-ST}} and \textbf{\textit{GT-Nudge}} treatments. Furthermore, while $74.9\%$ of our sample ($152$ of $203$ participants) report making patient-assignment decisions in their clinical role, we do not find decision-making differences between participants with and without this experience. 

In addition to the strategy chosen in the incentivized round, we examined behavior in Round 1, where participants make assignment decisions in real time. Batching rates in this round range from 58.4\% (\textbf{\textit{GT-Nudge}}) to 75.2\% (\textbf{\textit{IT}}), directionally consistent with the Round 2 results. Notably, batching rates \emph{increase} over the course of the practice round in all treatments: comparing the first half of decisions to the second half, batching rises by 10.6 (resp.: 14.0, 11.2, 1.0) percentage points in \textbf{\textit{IT}} (resp.: \textbf{\textit{GT}}, \textbf{\textit{GT-Nudge}}, \textbf{\textit{GT-ST}}), with the increase being significant in three of the four treatments (Signed-rank tests, $p = 0.021$, $p = 0.001$, and $p = 0.007$, $p = 0.933$). This suggests that after some initial experimentation participants' behavior converges towards batching, as they become more familiar with the task. 
Decisions in the practice round also respond to feedback. Among the 786 decisions in which a participant chose not to batch despite having the option, the next-decision batching rate is 10.1 percentage points higher when the case left behind was still unclaimed than when the partner had picked it up (26.5\% vs.\ 15.9\%, $p = 0.006$, see Table~\ref{tab:round1_regret} for the full regression). This suggests that even decision-makers with a low propensity to batch may learn this behavior as they observe the system evolve.

\subsection{Experiment 2: Emergency Medicine Physician Sample}\label{subsec:Physicians}
To verify that batching continues to persist for decision-makers with extensive patient self-assignment experience, we conducted a partial replication of our experiment with verified ED physicians.\footnote{The physician version of the experiment was shorter than the Prolific version, taking 7.2 minutes on average compared with 15.2 minutes for the Prolific sample. To reduce the time burden on physician participants, we omitted the practice round (Round 1); after going through the instructions, participants proceeded directly to the strategy-commitment round (Round 2).} In particular, we recruited practicing emergency medicine physicians through the EMDocs Facebook group, a gated online community of over 30{,}000 emergency physicians. Membership in the group requires verification of medical credentials. Participant recruitment (posting a link to the experiment on the group board) was done by one of the authors who is a practicing emergency physician and thus has access to the group. As pre-registered, physicians were assigned to either the \textbf{\textit{GT}} or the \textbf{\textit{GT-Nudge}} treatment in order to conserve subjects for the treatments most relevant to our research questions. 

Our initial target was 100 physician participants across the two treatments. A total of 85 unique physicians completed the experiment across multiple waves of recruitment. After excluding 2 physicians who reported technical difficulties and 10 physicians who did not complete the study, the analysis sample consists of $N = 73$ physicians: \textbf{\textit{GT}} ($n = 41$) and \textbf{\textit{GT-Nudge}} ($n = 32$). The mean age is 42.1 years (range 28--74); 41.1\% identify as female. The modal clinical experience category is 8--15 years (65.8\% of the sample), and nearly all physicians (97.8\%) report assigning patients to themselves in their clinical practice.

\subsubsection{Hypothesis Tests} 
Table~\ref{tab:physician_results} reports batching rates by treatment. Recall that the physician sample was assigned to two of the four treatments from Experiment~1, allowing us to re-test Hypothesis 1b (whether group throughput incentives reduce batching below 50\%) and Hypothesis 3a (whether the nudge reduces batching relative to group incentives alone). Under \textbf{\textit{GT}}, 73.2\% of physicians chose to batch, significantly above the 50\% benchmark (test of proportions, $p = 0.003$). Hypothesis 1b is not supported: even among practicing emergency physicians, a substantial majority batches despite group throughput incentives that reward the opposite behavior. Under \textbf{\textit{GT-Nudge}}, the batching rate falls to 56.2\%, corresponding to a 17 percentage-point reduction relative to \textbf{\textit{GT}}. This difference is in the predicted direction and marginally significant (one-sided test of proportions, $p = 0.065$). As in Experiment 1, we check for potential predictors of behavior in the physician sample (age, gender, clinical experience) using regression analysis but do not find any significant results.

\begin{table}[H]
\centering\footnotesize
\caption{Batching Behavior by Treatment (Physician Sample)}
\label{tab:physician_results}
\renewcommand{\arraystretch}{1.15}
\begin{tabular}{lcccc}
\hline
Treatment & $N$ & Never Batch & Always Batch & Batch Rate (\%) \\
\hline
\textbf{\textit{GT}}          & 41 & 11 & 30 & 73.2 \\
\textbf{\textit{GT-Nudge}} & 32 & 14 & 18 & 56.2 \\
\hline
Total & 73 & 25 & 48 & 65.8 \\
\hline
\end{tabular}
\end{table}

\begin{result}\label{res:physicians}
Physicians batch at rates that are higher than 50\% under group throughput incentives (Hypothesis 1b not supported). The nudge reduces batching by a magnitude similar to the Prolific sample (Hypothesis 3a marginally supported, $p = 0.065$).
\end{result}

\subsection{Discussion}
Comparing the Prolific and the physician samples, the physician results are largely consistent with the Prolific sample: batching rates are high under group incentives, and the nudge reduces batching by a similar magnitude. However, two differences stand out. First, physician batching rates are lower than Prolific batching rates in both treatments: 73.2\% vs.\ 94.1\% under \textbf{\textit{GT}} (test of proportions, $p = 0.013$) and 56.2\% vs.\ 76.9\% under \textbf{\textit{GT-Nudge}} ($p = 0.081$), suggesting that clinical experience reduces the bias but does not eliminate it.  Second, under \textbf{\textit{GT-Nudge}}, the physician batching rate of 56.2\% is close to the 50\% benchmark (test of proportions, $p = 0.596$), suggesting that the combination of clinical experience and explicit information about the externality of batching brings behavior close to the point of indifference between the two strategies.

Taken together, the experimental results provide evidence for what we term a \textit{personal productivity bias}: decision-makers systematically prefer larger self-assignments even when doing so reduces both system performance and their own earnings. This bias becomes stronger with learning and is not eliminated by experience: while more experienced participants in the Prolific sample batch less, and physicians batch less than the Prolific sample, a majority still batches under group incentives. These behaviors echo the field data in \textsection\ref{sec:FieldData} where batching was shown to be widespread across all five emergency departments. The experiments further clarify that batching is pervasive  whether or not it is aligned with the incentives. 

\subsubsection{Post-Experiment Survey}
Text-form survey responses reveal some of the reasoning behind batching. In particular, the responses suggest three dominant motives that participants use to explain their batching behavior (See \ref{app:quote_analysis} for details on the methodology and analysis). Two of the motives are consistent with the personal productivity bias: a \textit{busyness} motive, i.e., a preference for being busy or avoiding idle time, and a \textit{throughput} motive, i.e., the (incorrect) belief that batching maximizes the overall number of patients treated by the ED. The third is a \textit{habit} motive, in which participants mention their real-life behavior or experience (outside of the experiment) as the main reason for batching.\footnote{Busyness keywords include ``busy'', ``idle'', ``waste time'', ``back and forth'', ``lazy'', ``multitask''. Example: \textit{``I always prefer to take on more tasks at once to knock down the time spent wasted''}. Throughput keywords include ``more patients'', ``as many patients as possible'', ``maximize patients/treated'', ``treat more'', ``clear queue'', ``move quickly'', and ``benefit for the group''. Example: \textit{``Trying to maximize benefit for the group''}. Habit keywords include ``in real life'', ``what I do at work'', ``what I usually do'', ``I am used to'', ``based on how I practice'', and ``in all my years''. Example: \textit{``That's what I do in real life''}. See Appendix~\ref{app:quote_analysis} for the classification procedure and a detailed breakdown by treatment.}  Among 131 Prolific batchers in all treatments where batching was suboptimal (\textbf{\textit{GT}}, \textbf{\textit{GT-Nudge}}, and \textbf{\textit{GT-ST}}), $16.0\%$ give a busyness motive, $59.5\%$ give a throughput motive, and $2.3\%$ give a habit motive. Among 32 physician batchers who provided a description, the corresponding shares are $37.5\%$, $31.2\%$, and $15.6\%$. That is, Prolific participants predominantly believe that batching helps maximize total throughput, while physicians also cite busyness or habit. 

The throughput motive can be addressed via additional training or by making the externalities of batching more salient, as we have done in the \textbf{\textit{GT-Nudge}} treatment. Indeed, the share of participants mentioning the throughput motive drops from $60.8\%$ in \textbf{\textit{GT}} to $55.8\%$ in \textbf{\textit{GT-Nudge}} in the Prolific sample, and from $28.6\%$ to $15.8\%$ in the physician sample (Panel C of Table~\ref{tab:quote_classifications}). The busyness motive, however, is a more intrinsic preference that may be difficult to address through nudging. We next discuss how such a preference can be represented mathematically (in a utility function), and what this implies for future models of service design. 

\subsubsection{Implications for Service Design Models}\label{subsec:DiscussionModels}
Our conversations with ED management and physician groups indicate that they spend considerable time and effort on the design of physician compensation packages. A large analytical literature examines optimal contract design in these types of settings \citep[see, for example,][]{GilbertWeng1998Incentive, CachonZhang2007FastService, HoppIravaniYuen2007}. The standard approach is to model server utility as a function of monetary compensation, sometimes net of an effort cost. \citet{ShumskyPinker2003} include a nonmonetary parameter $h$ for the server's preference to spend (or avoid) time on treatment, and use it to study compensation systems that align gatekeeper behavior with system-optimal referral rates. Our experimental data suggest that in the emergency department setting, $h$ is positive and large enough that financial incentives alone do not align individually optimal and socially desirable decisions. A simple way to capture this analytically is to add a busyness term to the server's utility of the form $U_i = \text{wage}(\cdot) + h \cdot \mathbf{1}\{\text{busy}\}$ with $h > 0$. Under this objective, aligning worker behavior with the social optimum likely requires going beyond standard compensation contracts and may include information design or  compensation contracts that respond to the current state of the service system.

The personal productivity bias suggested in our data may also play a role in models of optimal queueing control. The classic pooled-vs-dedicated literature \citep[e.g.,][]{MandelbaumReiman1998Pooling, Whitt1999Partitioning, vanDijkvanderSluis2009Pooling} compares either a single shared queue or dedicated per-server queues, typically with exogenous routing of patients to physicians. However, service systems in healthcare are often hybrid. A shared pool of unassigned cases may feed into each server's personal caseload, and the rate at which cases move from the shared pool to a personal queue is itself an endogenous server choice. Several behaviorally inspired models have examined related behaviors, for example, weakened customer ownership in pooled systems \citep{ArmonyRoelsSong2021} and customer-side over-joining of pooled queues \citep{SunarTuZiya2021}.\footnote{See also \cite{SongTuckerMurrell2015,SongArmonyRoels2024} for empirical work that supports richer models of server behavior in queueing systems.} These papers show that accounting for behavioral preferences can change prescriptions for optimal service design. In the same vein, adding a server-side preference for being busy ($h \cdot \mathbf{1}\{\text{busy}\}$) may change prescriptions for optimal queue configuration in systems where servers self-assign from a shared pool.

\subsubsection{Implications for Practice}
In addition to informing the service operations literature, our results have  implications for managing patient assignment in practice. First, the pervasiveness of batching under all incentive systems across the five physician groups suggests shifting from individual to group throughput metrics is likely not sufficient to address the problem. Second, the effectiveness of the nudge in \textbf{\textit{GT-Nudge}} suggests that making the consequences of batching visible at the point of decision can reduce the bias. In practice, this could be implemented through the electronic health record interface: for example, displaying partner idle time or flagging when self-assigning a second patient would leave a colleague without work. Third, and more broadly, our results suggest that the default design of ED tracking boards (see Fig. 1 for an example), which allows physicians to freely select and reserve multiple cases, may facilitate a behavior that is harmful to system performance. Adding frictions, such as constraining the interface to allow only single-patient assignment unless explicit justification is given for batch-assignment, could help mitigate the bias. Similar hurdles have been shown to effectively counteract other biases in healthcare \citep[e.g.,][]{patel2018nudge,kim2020admission}, and we believe that  more work needs to be done to identify how changes to software interfaces in medical decision-making can help improve system performance.

\section{Concluding Remarks}\label{sec:Conclusion}

This paper studies physician choice of patient assignment in emergency departments and asks why physicians frequently claim multiple patients at once even though doing so can slow down patient flow. The field data show that batching is widespread and common at every hospital in our sample and that it leads to a 9\% to 12\% increase in LOS for the batched patients, even after controlling for physician fixed effects, clinical acuity, demographics, and ED congestion. We then show analytically that while personal-throughput maximization can make batching individually optimal under piece-rate compensation, assigning exactly one patient at a time minimizes total patient sojourn time and maximizes throughput. Our experiments then examine whether human decision-makers respond to incentives in the way that the model predicts. Our main finding is that they do not: under group throughput incentives, where not batching maximizes both system performance and the participant's own bonus, 94\% of healthcare workers and 73\% of practicing emergency physicians still choose to batch. We term this behavior a \textit{personal productivity bias}: a preference for the strategy that keeps the decision-maker busier in the short term, even when it costs the system and the decision-maker themselves in the long term. A one-sentence informational nudge that highlights the consequence of batching for partner idleness reduces batching by 17 percentage points in both samples.

Stepping back, our findings raise a broader question about how operations researchers and managers should think about incentives in service and healthcare settings. The service operations literature often treats incentive misalignment as the canonical reason for system-suboptimal behavior, and proposes incentive realignment as the canonical fix \citep[e.g.,][]{GilbertWeng1998Incentive, CachonZhang2007FastService}. The personal productivity bias suggests that this view may be incomplete when people draw from a shared workload. Decision-makers in our setting prefer larger personal workloads not because they are strategically free-riding (the partner's behavior is fixed and is common knowledge) and not because they misunderstand the incentives, but because being busy feels productive and being idle feels wasteful, and batch-assignment increases short term busyness. Our results therefore suggest that appropriate training and the design of the  interface may matter as much (or more) as the design of the compensation contract.

While we have focused on the negative consequences of batching, it is worth emphasizing that there are cases when batching may be beneficial. From the individual perspective, batching can be optimal under piece-rate compensation, as in our \textbf{\textit{IT}} treatment. From the system perspective, batching can improve efficiency in settings where switching from patient care to administrative work has a fixed cost and combining multiple cases helps spread that cost \citep[as in][]{feizi2023batch}. Batching can prevent the queue of unassigned cases from growing beyond a certain threshold, which is known to add stress to the system and can degrade the performance of physicians and nursing staff \citep{Litvak2005Variability, KCTerwiesch2009, BerryJaekerTucker2017}. Psychologically, batching may also be helpful from the patients' perspective as it moves them from a passive, unassigned waiting state into active treatment, which can improve perceived progress toward service completion \citep[see, e.g.,][for related psychological mechanisms]{SomanShi2003}.

Our work has the following limitations. First, the field analysis is observational and is at the patient level: we identify an increase in LOS for batched patients but we are not able to identify the net effect of batching on overall ED throughput. Second, the analytical model assumes homogeneous physicians, exogenous partner behavior, exponential service times, and patient blocking under finite room capacity; relaxing any of these assumptions to allow acuity-dependent service times, heterogeneous physician speed, or strategic interaction between physicians would make for a richer characterization of when batching can or cannot be rationalized as system-optimal. Third, our experiments are stylized, with a computerized partner and short horizons. A richer experiment design would include learning and team dynamics. More generally, while our results replicate across a healthcare-worker sample and a verified emergency-physician sample, the personal productivity bias is unlikely to be confined to physicians, and understanding how it interacts with team composition, supervision, and incentives in other service systems may be a promising direction.

\bibliographystyle{informs2014}
\bibliography{support_files/lit}

\ECSwitch
\ECHead{Electronic Companion}
\small

\medskip
\noindent\textbf{Contents}
\begin{itemize}[leftmargin=2em, itemsep=1pt, topsep=2pt]
\item \hyperref[app:empirics]{EC.1\quad Empirical Appendix}
\begin{itemize}[leftmargin=2em, itemsep=1pt, topsep=1pt]
  \item \hyperref[app:empirics:busyness]{EC.1.1\quad Variation in ED Busyness Across Hospitals}
  \item \hyperref[app:empirics:logit]{EC.1.2\quad Conditional Logit Analysis of Batching Decisions}
  \item \hyperref[app:empirics:regression]{EC.1.3\quad Linear Regression Results with Standard Errors}
\end{itemize}
\item \hyperref[app:proofs]{EC.2\quad Proofs}
\item \hyperref[app:experiment]{EC.3\quad Experimental Appendix}
\begin{itemize}[leftmargin=2em, itemsep=1pt, topsep=1pt]
  \item \hyperref[app:experiment:exclusion]{EC.3.1\quad Pre-Registration and Exclusion Criteria}
  \item \hyperref[app:experiment:incentives]{EC.3.2\quad Incentives in each Treatment}
  \item \hyperref[app:demographics]{EC.3.3\quad Additional Experimental Results}
  \item \hyperref[app:instructions]{EC.3.4\quad Experiment Instructions}
  \item \hyperref[app:screenshots]{EC.3.5\quad Additional Experiment Interface Screenshots}
\end{itemize}
\end{itemize}
\medskip

\section{Empirical Appendix}\label{app:empirics}

\subsection{Variation in ED Busyness Across Hospitals}\label{app:empirics:busyness}

\begin{figure}[h!]
\centering
\caption{Distribution of ED Busyness by Hospital}
\label{fig:busyness_distribution}
\includegraphics[width=0.95\textwidth]{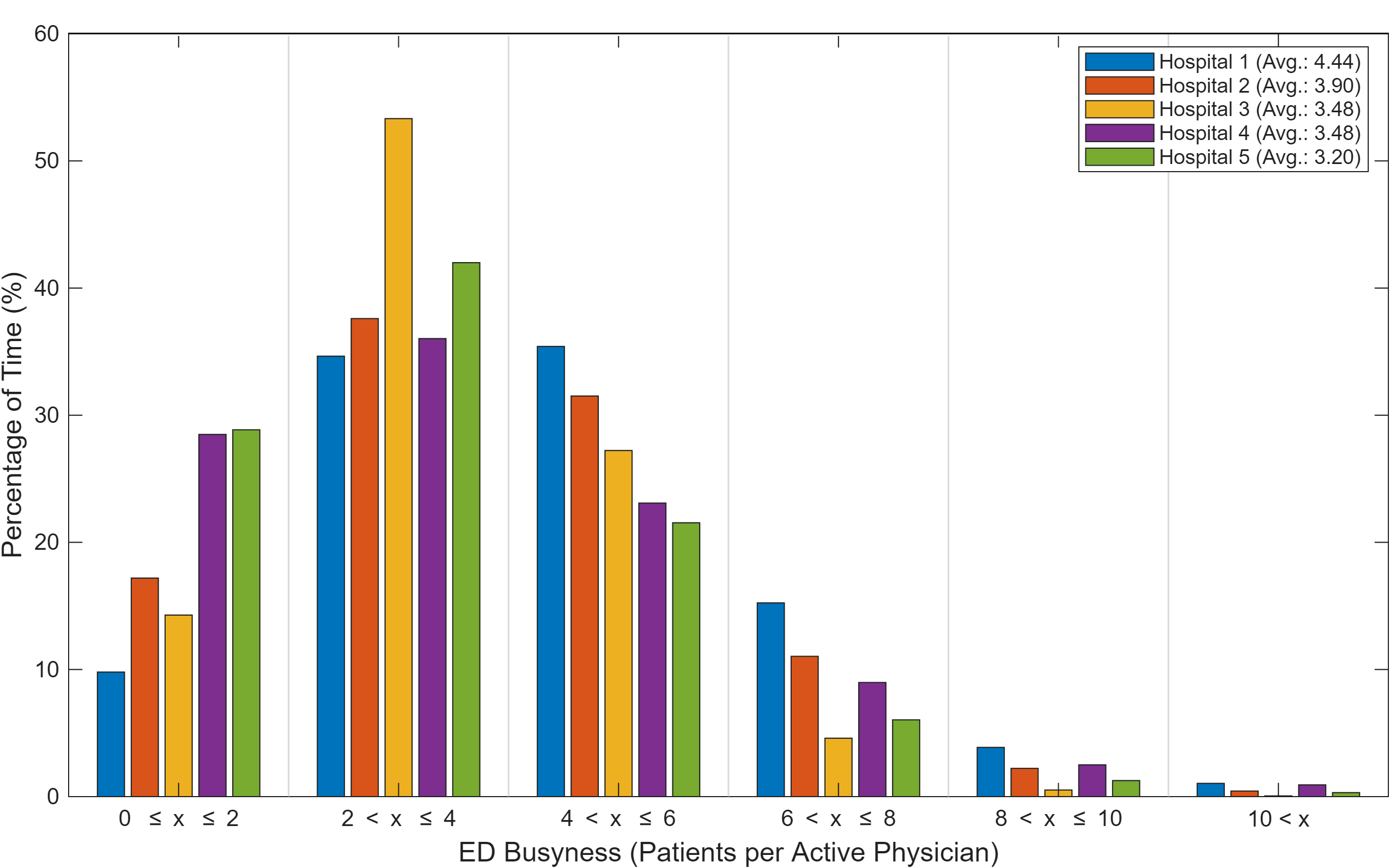}
\end{figure}

A key identifying assumption underlying our regression analysis is that there is meaningful variation in ED congestion across time and hospitals. Figure~\ref{fig:busyness_distribution} documents this variation by plotting the distribution of ED busyness (defined as the ratio of total active cases to the number of physicians on shift) separately for each of the five hospitals in our sample. Two observations are in order. First, there is substantial within-hospital variation in busyness over time, with the distribution spread across a wide range of values at every hospital. This within-hospital variation is the primary source of identifying variation for the busyness coefficient in our regression. Second, there is meaningful between-hospital variation in the average level of busyness, ranging from a weighted average of 3.20 at Hospital 5 to 4.44 at Hospital 1. That is, the EDs in our sample operate in a regime where the queue is neither empty nor severely overloaded most of the time, with each physician typically managing three to five active cases at any given moment. Furthermore, two of the hospitals (Hospital 4 and 5) operate at a relatively light load (0 to 2 patients per physician) for about 30\% of the time. Together, these observations confirm that busyness varies sufficiently across both time and hospitals to support the identification strategy used throughout \textsection\ref{sec:FieldData}.

\subsection{Conditional Logit Analysis of Batching Decisions}\label{app:empirics:logit}

To complement the model-free evidence in \textsection\ref{sec:field:model-free} and provide a more direct test of load-adaptive batching behavior, we estimate a conditional logit model of the batching decision at the level of the individual batching opportunity. Recall from \S\ref{sec:field:model-free} that a batching opportunity is defined as an instance in which a physician makes an assignment and at least one additional unassigned patient remains in the ED. The two key covariates are the focal physician's census at the moment of the initial assignment in a round (the number of cases they are currently managing, measured immediately after their first assignment, which reflects their baseline workload at the moment the batching decision is made), and the total number of unassigned patients in the ED at that moment.

We estimate the model using a conditional logit (fixed-effects logit) specification, conditioning on physician-level sufficient statistics to absorb physician fixed effects without the incidental parameters bias that would arise from including physician dummies in a nonlinear model. For the pooled specification across all five hospitals, we additionally include hospital indicator variables as standard regressors, with Hospital 1 as the reference category. All specifications include controls for physician utilization rate over the prior eight hours, physician cumulative experience, and seasonality fixed effects (year, month, weekday, hour). The model is estimated using the \texttt{clogit} function in R with the Efron approximation for tied event times.

Table~\ref{tab:conditional_logit} reports the estimated coefficients for the two focal covariates across the pooled and individual hospital specifications. The results strongly validate the load-adaptive patterns documented in Table~\ref{tab:batching_by_census_unassigned}. The coefficient on focal physician census is negative and highly significant in all six specifications, confirming that even after conditioning on each physician's underlying propensity to batch through the fixed effect, physicians respond to their own workload: a physician who has just taken on their first case in a round is substantially more likely to batch than one who already has several cases underway. This suggests that batching is used as a deliberate workload management tool, with physicians self-regulating their personal caseload in real time. The coefficient on patients unassigned is positive and highly significant in all six specifications, confirming that physicians are more likely to batch when the shared pool of unassigned cases is large. Together, these two patterns suggest that even after correcting for each physician's baseline tendency to batch through the fixed effect, physicians at large respond to the state of the system in a way that appears to reflect a perceived need to load-balance: they claim more cases when there are more cases to claim and fewer cases when they are already stretched.

\begin{table}[h!]
\centering\footnotesize
\caption{Conditional Logit Results: Probability of Batching at a Batching Opportunity}
\label{tab:conditional_logit}
\begin{tabular}{lc|ccccc}
\hline
Variable & All Hospitals & Hospital 1 & Hospital 2 & Hospital 3 & Hospital 4 & Hospital 5 \\
\hline
\textit{Focal Physician Census}
    & $-$0.0930*** & $-$0.1101*** & $-$0.1242*** & $-$0.0717*** & $-$0.0650*** & $-$0.1478*** \\
    & (0.0010)     & (0.0017)     & (0.0025)     & (0.0020)     & (0.0026)     & (0.0033)     \\[4pt]
\textit{Patients Unassigned}
    & 0.0791***    & 0.0712***    & 0.1269***    & 0.0584***    & 0.1365***    & 0.1503***    \\
    & (0.0005)     & (0.0010)     & (0.0017)     & (0.0008)     & (0.0027)     & (0.0034)     \\
\hline
Hospital Fixed Effects      & \checkmark &            &            &            &            &            \\
Physician Fixed Effects     & \checkmark & \checkmark & \checkmark & \checkmark & \checkmark & \checkmark \\
Physician Experience        & \checkmark & \checkmark & \checkmark & \checkmark & \checkmark & \checkmark \\
Physician Utilization       & \checkmark & \checkmark & \checkmark & \checkmark & \checkmark & \checkmark \\
Seasonality                 & \checkmark & \checkmark & \checkmark & \checkmark & \checkmark & \checkmark \\
\hline
$N$                         & 1{,}042{,}699 & 264{,}127 & 207{,}792 & 382{,}192 & 106{,}789 & 81{,}765 \\
Log-Likelihood              & $-$2{,}616{,}536.6 & $-$808{,}942.7 & $-$449{,}952.4 & $-$767{,}917.8 & $-$304{,}849.7 & $-$266{,}522.8 \\
\hline
\end{tabular}
\\[4pt]
\begin{minipage}{0.88\textwidth}
\scriptsize
\textit{Notes.} *** $p < 0.001$. Standard errors in parentheses. Coefficients are log-odds. The dependent variable is an indicator equal to one if the physician's next assignment occurs within five minutes of the current assignment. The model is estimated as a conditional logit conditioning on physician-level sufficient statistics to absorb physician fixed effects without incidental parameters bias. The All Hospitals specification additionally includes hospital indicator variables with Hospital 1 as the reference category. All specifications include controls for physician utilization rate over the prior eight hours, cumulative physician experience, and seasonality fixed effects (year, month, weekday, hour of day).
\end{minipage}
\end{table}

\subsection{Linear Regression Results with Standard Errors}\label{app:empirics:regression}

Table~\ref{tab:regression_results_se} reproduces the regression results from Table~\ref{tab:regression_results} in the main text with standard errors reported in parentheses below each coefficient estimate, providing a more complete picture of estimation precision. All specifications, variable definitions, and controls are identical to those described in \textsection\ref{sec:field:reg}.

\begin{table}[h!]
\centering\footnotesize
\caption{Linear Regression Results with Standard Errors: Length of Stay (Minutes)}
\label{tab:regression_results_se}
\begin{tabular}{lc|ccccc}
\hline
Variable & All Hospitals & Hospital 1 & Hospital 2 & Hospital 3 & Hospital 4 & Hospital 5 \\
\hline
\textit{Batch\_2}
    & 7.46***  & 8.86***  & 5.68***  & 7.61***  & 7.83***  & 5.22***  \\
    & (0.17)  & (0.32)  & (0.34)  & (0.36)  & (0.43)  & (0.41)  \\[4pt]
\textit{Batch\_3}
    & 14.77*** & 16.14*** & 13.20*** & 15.05*** & 13.91*** & 13.23*** \\
    & (0.33)  & (0.57)  & (0.76)  & (0.78)  & (0.83)  & (0.76)  \\[4pt]
\textit{Batch\_4}
    & 22.87*** & 26.00*** & 21.93*** & 20.58*** & 26.10*** & 16.97*** \\
    & (0.63)  & (0.98)  & (1.51)  & (1.77)  & (1.67)  & (1.33)  \\[4pt]
\textit{Batch\_5+}
    & 30.45*** & 35.13*** & 31.78*** & 29.39*** & 36.03*** & 19.11*** \\
    & (0.97)  & (1.36)  & (2.53)  & (3.57)  & (2.70)  & (1.95)  \\[4pt]
\textit{Focal Physician Census}
    & 0.88***  & 1.01***  & 0.39***  & 0.68***  & 1.98***  & 1.78***  \\
    & (0.04)  & (0.08)  & (0.08)  & (0.09)  & (0.11)  & (0.13)  \\[4pt]
\textit{ED Busyness}
    & 6.65***  & 6.45***  & 5.99***  & 10.65*** & 4.70***  & 4.98***  \\
    & (0.05)  & (0.09)  & (0.10)  & (0.12)  & (0.12)  & (0.14)  \\[4pt]
\textit{Handoff}
    & 16.77*** & 20.24*** & 50.81*** & 10.05*** & $-$3.15*** & 10.38*** \\
    & (0.35)  & (0.72)  & (0.79)  & (0.65)  & (0.90)  & (0.98)  \\[4pt]
\textit{Discharged}
    & 15.95*** & 6.65***  & 14.22*** & 20.28*** & 17.43*** & 14.67*** \\
    & (0.17)  & (0.36)  & (0.34)  & (0.35)  & (0.44)  & (0.44)  \\
\hline
Hospital Fixed Effects      & \checkmark &            &            &            &            &            \\
Physician Fixed Effects     & \checkmark & \checkmark & \checkmark & \checkmark & \checkmark & \checkmark \\
Physician Experience        & \checkmark & \checkmark & \checkmark & \checkmark & \checkmark & \checkmark \\
Physician Utilization       & \checkmark & \checkmark & \checkmark & \checkmark & \checkmark & \checkmark \\
Seasonality                 & \checkmark & \checkmark & \checkmark & \checkmark & \checkmark & \checkmark \\
Patient Demographics        & \checkmark & \checkmark & \checkmark & \checkmark & \checkmark & \checkmark \\
Patient Prior Visits        & \checkmark & \checkmark & \checkmark & \checkmark & \checkmark & \checkmark \\
Chief Complaint             & \checkmark & \checkmark & \checkmark & \checkmark & \checkmark & \checkmark \\
Acuity Level                & \checkmark & \checkmark & \checkmark & \checkmark & \checkmark & \checkmark \\
Testing/Medicine Intensity  & \checkmark & \checkmark & \checkmark & \checkmark & \checkmark & \checkmark \\
\hline
$N$
    & 1{,}442{,}955 & 342{,}379 & 296{,}458 & 439{,}335 & 184{,}662 & 180{,}121 \\
$R^2$
    & 0.3553 & 0.3582 & 0.3948 & 0.3020 & 0.3517 & 0.3729 \\
Adjusted $R^2$
    & 0.3549 & 0.3575 & 0.3941 & 0.3013 & 0.3507 & 0.3718 \\
\textbf{Weighted Avg \% Effect on LOS}
    & \textbf{10.09\%} & \textbf{12.21\%} & \textbf{8.78\%} & \textbf{8.75\%} & \textbf{10.36\%} & \textbf{8.94\%} \\
\hline
\end{tabular}
\\[4pt]
\begin{minipage}{0.88\textwidth}
\scriptsize
\textit{Notes.} *** $p < 0.001$. Standard errors in parentheses. Regression coefficients are reported in minutes. The All Hospitals specification additionally includes hospital indicator variables. While physician IDs are unique across hospitals, 47 physicians treated cases at more than one hospital in the sample; of these, only 9 treated 500 or more cases at two or more hospitals. These cross-hospital physicians provide the identifying variation for the hospital fixed effects in the pooled specification.
\end{minipage}
\end{table}

\section{Proofs} \label{app:proofs}

\subsection*{Proof of Proposition 1}

Throughout, we work with the experimental parameters: two physicians (a focal player and a partner), $M = 4$ treatment rooms, groups of three patients arriving at rate $\lambda$, and exponential service times with rate $\mu$. Each state is represented by the triple $s = (U, A_1, A_2)$, where $U$ is the number of unassigned patients, $A_1 \in \{0,1,2\}$ is the number of cases assigned to the focal physician, and $A_2 \in \{0,1\}$ indicates whether the partner is active. The reward rate for personal throughput is $\rho(s) = \mu \cdot \mathbf{1}[A_1 \geq 1]$, counting only completions by the focal physician.

\subsubsection*{Step 1: Generator Matrices}

Because assignment decisions are instantaneous, certain states are never occupied for a positive duration and therefore need not appear as distinct states in the CTMC. Instead, they collapse into direct transitions between stable states. To see this, consider the three situations in which the player faces a genuine choice between claiming one or two patients.

First, when the system is in state $(0,0,0)$ and a group of three patients arrives, the system passes instantaneously through state $(3,0,0)$. The partner immediately claims one patient, bringing the system to the intermediate state $(2,0,1)$, at which point the player claims either one patient, leading to state $(1,1,1)$, or two patients, leading to state $(0,2,1)$. Second, when the system is in state $(0,0,1)$ and a group of three patients arrives, the system passes instantaneously through $(3,0,1)$, and the player immediately claims either one patient, leading to state $(2,1,1)$, or two patients, leading to state $(1,2,1)$. Third, when the system is in state $(2,1,1)$ and the player completes treatment, the system passes instantaneously through $(2,0,1)$, at which point the player claims either one patient, leading to state $(1,1,1)$, or two patients, leading to state $(0,2,1)$. In all three cases, the intermediate states are transient and are collapsed into direct transitions in the generator matrix. This collapsing also clarifies the structure of the Poisson equations in Step 2: the relative value comparison at each decision epoch reduces to comparing the value of the state the system enters under batching against the value of the state it enters under no-batching.

There are nine reachable states under the experimental parameters: $\mathcal{S} = \{(0,0,0),\,(0,0,1),\,(0,1,0),\,(0,2,0),\,(0,1,1),\,(0,2,1),\,(1,1,1),\,(1,2,1),\,(2,1,1)\}$; the policies differ only in the generator matrices, which reflect the different post-decision states entered at each decision epoch. Under policy B the generator is
\[
\scriptsize
\boldsymbol{Q}_{\mathrm{B}} =
\renewcommand{\arraystretch}{1.15}
\begin{array}{c|ccccccccc}
& (0,0,0) & (0,0,1) & (0,1,0) & (0,2,0) & (0,1,1) & (0,2,1) & (1,1,1) & (1,2,1) & (2,1,1) \\
\hline
(0,0,0) & -\lambda & & & & & \lambda & & & \\
(0,0,1) & \mu & -(\lambda+\mu) & & & & & & \lambda & \\
(0,1,0) & \mu & & -(\lambda+\mu) & & & & & & \lambda \\
(0,2,0) & & & \mu & -(\lambda+\mu) & & & & \lambda & \\
(0,1,1) & & \mu & \mu & & -(\lambda+2\mu) & & & & \lambda \\
(0,2,1) & & & & \mu & \mu & -(\lambda+2\mu) & & \lambda & \\
(1,1,1) & & & & & 2\mu & & -(\lambda+2\mu) & & \lambda \\
(1,2,1) & & & & & & \mu & \mu & -2\mu & \\
(2,1,1) & & & & & & \mu & \mu & & -2\mu
\end{array}
\]

Arrivals occur at rate $\lambda$ and service completions occur at rate $\mu$ per active physician. A similar thought process leads to the generator matrix for policy NB. The two matrices are identical except in rows $(0,0,0)$, $(0,0,1)$, and $(2,1,1)$, reflecting the fact that under NB the player always claims exactly one patient at each decision epoch rather than two.
\[
\scriptsize
\boldsymbol{Q}_{\mathrm{NB}} =
\renewcommand{\arraystretch}{1.15}
\begin{array}{c|ccccccccc}
& (0,0,0) & (0,0,1) & (0,1,0) & (0,2,0) & (0,1,1) & (0,2,1) & (1,1,1) & (1,2,1) & (2,1,1) \\
\hline
(0,0,0) & -\lambda & & & & & & & & \lambda \\
(0,0,1) & \mu & -(\lambda+\mu) & & & & & & & \lambda \\
(0,1,0) & \mu & & -(\lambda+\mu) & & & & & & \lambda \\
(0,2,0) & & & \mu & -(\lambda+\mu) & & & & \lambda & \\
(0,1,1) & & \mu & \mu & & -(\lambda+2\mu) & & & & \lambda \\
(0,2,1) & & & & \mu & \mu & -(\lambda+2\mu) & & \lambda & \\
(1,1,1) & & & & & 2\mu & & -(\lambda+2\mu) & & \lambda \\
(1,2,1) & & & & & & \mu & \mu & -2\mu & \\
(2,1,1) & & & & & & & 2\mu & & -2\mu
\end{array}
\]

\subsubsection*{Step 2: Poisson Equations and Optimality}

For each policy $k \in \{\mathrm{NB}, \mathrm{B}\}$, the relative value function $h_k(s)$ and long-run average personal throughput rate $g_k$ satisfy the Poisson equations
\begin{equation}\label{eq:poisson}
g_k + h_k(s) = \rho(s) + \sum_{s' \in \mathcal{S}^k} Q_k(s,s')\,h_k(s'), \qquad \forall\, s \in \mathcal{S}^k,
\end{equation}
with normalization $h_k(0,0,0) = 0$. The relative value $h_k(s)$ measures the expected cumulative personal throughput advantage of starting in state $s$ rather than in the empty and idle state $(0,0,0)$, with the steady-state average rate $g_k$ subtracted out. For each policy, the system \eqref{eq:poisson} yields a linear system in the unknowns $\{h_k(s) : s \in \mathcal{S}^k\}$ and $g_k$, which is solved in closed form subject to the normalization constraint.

Batching is optimal at every decision epoch if and only if the relative value of the post-batch state weakly exceeds that of the post-no-batch state at each of the three decision epochs identified in Step 1. At the decision epochs arising from states $(0,0,0)$ and $(2,1,1)$, the relevant comparison is $h_k(0,2,1) \geq h_k(1,1,1)$. At the decision epoch arising from state $(0,0,1)$, the relevant comparison is $h_k(1,2,1) \geq h_k(2,1,1)$. Solving the linear system \eqref{eq:poisson} and computing these differences yields, under policy B,
\begin{equation}
\Delta^1_{\mathrm{B}} \equiv h_{\mathrm{B}}(0,2,1) - h_{\mathrm{B}}(1,1,1) = \frac{\mu^3\bigl(\lambda^4 + 7\lambda^3\mu + 18\lambda^2\mu^2 + 18\lambda\mu^3 + 8\mu^4\bigr)}{(\lambda+\mu)^2\bigl(\lambda^5 + 9\lambda^4\mu + 32\lambda^3\mu^2 + 53\lambda^2\mu^3 + 42\lambda\mu^4 + 16\mu^5\bigr)},
\end{equation}
\begin{equation}
\Delta^2_{\mathrm{B}} \equiv h_{\mathrm{B}}(1,2,1) - h_{\mathrm{B}}(2,1,1) = 0,
\end{equation}
and under policy NB,
\begin{equation}
\Delta^1_{\mathrm{NB}} \equiv h_{\mathrm{NB}}(0,2,1) - h_{\mathrm{NB}}(1,1,1) = \frac{2\mu^3\bigl(\lambda^4 + 7\lambda^3\mu + 18\lambda^2\mu^2 + 18\lambda\mu^3 + 8\mu^4\bigr)}{(\lambda+2\mu)^2\bigl(\lambda^5 + 8\lambda^4\mu + 25\lambda^3\mu^2 + 34\lambda^2\mu^3 + 24\lambda\mu^4 + 8\mu^5\bigr)},
\end{equation}
\begin{equation}
\Delta^2_{\mathrm{NB}} \equiv h_{\mathrm{NB}}(1,2,1) - h_{\mathrm{NB}}(2,1,1) = \frac{\mu^3\bigl(\lambda^4 + 7\lambda^3\mu + 18\lambda^2\mu^2 + 18\lambda\mu^3 + 8\mu^4\bigr)}{(\lambda+2\mu)^2\bigl(\lambda^5 + 8\lambda^4\mu + 25\lambda^3\mu^2 + 34\lambda^2\mu^3 + 24\lambda\mu^4 + 8\mu^5\bigr)}.
\end{equation}
Since every coefficient in the numerator and denominator polynomials of $\Delta^1_{\mathrm{B}}$, $\Delta^1_{\mathrm{NB}}$, and $\Delta^2_{\mathrm{NB}}$ is strictly positive, each is strictly positive for all $\lambda, \mu > 0$, and $\Delta^2_{\mathrm{B}} = 0$ establishes weak optimality at the third decision epoch. Batching therefore weakly dominates no-batching at every decision epoch and is strictly optimal at the decision epochs arising from states $(0,0,0)$ and $(2,1,1)$, confirming that batching is the unique optimal policy for the focal physician. \qed

\subsection*{Proof of Proposition 2}

\textbf{Notation and assumptions.} Fix a realization of the arrival process, including arrival epochs and group sizes. Fix also an infinite sequence of potential service requirements $\{S_k\}_{k\geq 1}$, where $S_k$ is the service time assigned to the $k$-th service initiation. System parameters $\mathbf{\Theta} = \{M, N, \lambda, P, \mu\}$ are fixed. Physicians are homogeneous and follow non-idling, stationary assignment rules. Let $\mathcal{D}$ denote the class of admissible joint physician strategies, and define the single-case assignment policy $\mathbf{D}^\ast \in \mathcal{D}$ by $d_i(U, \mathbf{A}) \equiv 1$ for all $i, U, \mathbf{A}$. Let $\mathbf{D} \in \mathcal{D}$ be any other admissible joint policy, possibly assigning more than one patient at some decision epochs. For any policy $\Pi$ and time $t \geq 0$, let $N_\Pi(t)$ denote the number of patients in the system at time $t$, and let $C_\Pi(t)$ denote the cumulative number of completed services by time $t$.

We prove that on every fixed sample path and for all $t \geq 0$, $N_{\mathbf{D}^\ast}(t) \leq N_{\mathbf{D}}(t)$ and $C_{\mathbf{D}^\ast}(t) \geq C_{\mathbf{D}}(t)$. Consequently, the completion time and sojourn time of every admitted patient under $\mathbf{D}^\ast$ are no later than under $\mathbf{D}$, and $\mathbf{D}^\ast$ achieves weakly higher throughput over every finite horizon and in steady state.

We compare the evolution of the system under $\mathbf{D}^\ast$ and $\mathbf{D}$ on the fixed sample path. Up to the first assignment epoch at which the two policies differ, the systems evolve identically. Let $\tau$ denote the first decision epoch at which the policies diverge. At time $\tau$, a focal physician assigns more than one patient under policy $\mathbf{D}$, whereas under $\mathbf{D}^\ast$ the same physician assigns exactly one patient. By construction, system states immediately prior to $\tau$ are identical under both policies.

We first establish that service initiation for the second patient assigned under policy $\mathbf{D}$ at epoch $\tau$ occurs weakly later than under policy $\mathbf{D}^\ast$. If at time $\tau$ there exists at least one partner physician who is idle, then under $\mathbf{D}^\ast$ that physician would instead have been assigned this second patient and would have begun service immediately at time $\tau$. Under policy $\mathbf{D}$, the patient remains assigned to the focal physician and therefore cannot begin service at time $\tau$, so service initiation is strictly later under $\mathbf{D}$ in this case. If no partner physician is idle at time $\tau$, two subcases arise. If the focal physician completes service of the first assigned patient before any partner becomes idle, then under $\mathbf{D}^\ast$ the focal physician assigns the second patient upon returning to the terminal and begins service at exactly the same time as under $\mathbf{D}$. If instead a partner physician becomes idle before the focal physician completes service of the first patient, then under $\mathbf{D}^\ast$ that partner begins service on the second patient at the moment they become idle, whereas under $\mathbf{D}$ the patient remains assigned to the focal physician and begins service no earlier. In all cases, therefore, the service initiation time of the second patient under $\mathbf{D}^\ast$ is weakly earlier than under $\mathbf{D}$, with strict inequality in at least one feasible scenario.

Since service requirements are drawn from the fixed sequence $\{S_k\}$ in order of service initiation, and since at least one service initiation occurs strictly earlier under $\mathbf{D}^\ast$, the corresponding service completion also occurs strictly earlier. No service completion under $\mathbf{D}$ can occur earlier than its counterpart under $\mathbf{D}^\ast$, so all completion times under $\mathbf{D}^\ast$ are weakly earlier than under $\mathbf{D}$, with at least one strictly earlier. It follows that for all $t \geq \tau$, $N_{\mathbf{D}^\ast}(t) \leq N_{\mathbf{D}}(t)$. At every arrival epoch after $\tau$, the number of available treatment rooms under $\mathbf{D}^\ast$ is therefore weakly greater than under $\mathbf{D}$. Because admissions are monotone in available capacity, any patient blocked under $\mathbf{D}^\ast$ is also blocked under $\mathbf{D}$, while the converse need not hold, so $\mathbf{D}^\ast$ admits weakly more patients and achieves weakly higher throughput $\Lambda(\mathbf{\Theta}, \mathbf{D}^\ast) \geq \Lambda(\mathbf{\Theta}, \mathbf{D})$ over any finite horizon. Finally, earlier service initiation and earlier completion for every admitted patient imply that the sojourn time of each admitted patient under $\mathbf{D}^\ast$ is no greater than under $\mathbf{D}$, pointwise on the sample path. Taking expectations, or steady-state averages which exist because the state space is finite under blocking, yields $\mathbb{E}[W \mid \mathbf{\Theta}, \mathbf{D}^\ast] \leq \mathbb{E}[W \mid \mathbf{\Theta}, \mathbf{D}]$, completing the proof. \qed

\section{Experimental Details}\label{app:experiment}

\subsection{Pre-Registration and Exclusion Criteria}\label{app:experiment:exclusion}
The experimental design, hypotheses, sample sizes, and analysis plan were pre-registered on AsPredicted prior to data collection. The Prolific sample was pre-registered under AsPredicted \#269,600 and the physician sample under AsPredicted \#271,778.

We apply three pre-registered exclusion criteria to the Prolific sample:
\begin{enumerate}
\item \textbf{Completion time.} We exclude participants whose total experiment duration (from first page load to final submission) exceeds three times the sample median. The median completion time is 829 seconds (approximately 14 minutes), yielding a threshold of 2{,}487 seconds. Three participants are excluded under this criterion.
\item \textbf{Comprehension quiz errors.} Participants who make three or more total errors across the five comprehension quiz questions are excluded. Twenty-one participants are excluded under this criterion.
\item \textbf{Technical errors.} Participants who self-report technical difficulties that prevented them from completing the task as intended are excluded. Two participants are excluded under this criterion.
\end{enumerate}
No participant triggers more than one exclusion criterion. After exclusions, the Prolific analysis sample consists of $N = 203$ participants.

For the physician sample, the comprehension quiz and practice round were omitted to reduce the time burden. We exclude 2 physicians who reported technical difficulties, and 10 physicians who did not complete the study, yielding an analysis sample of $N = 73$.

\subsection{Incentives in each Treatment}\label{app:experiment:incentives}
Participants received a base participation fee of \$2.00 plus a performance bonus based on their performance in the incentivized round (Round 2). In the \textbf{\textit{IT}} treatment, the bonus was \$0.24 for each patient treated by the participant beyond a threshold of 6 patients. In \textbf{\textit{GT} }and \textbf{\textit{GT-Nudge}}, the bonus was \$0.60 for each patient treated by either the participant or the computerized partner beyond a joint threshold of 36 patients. In GT-ST, the bonus was \$0.014 for each time unit by which total patient sojourn time (waiting plus treatment) fell below a target of 1,200 time units.

Thresholds and per-unit payments were selected to equalize both the level of expected earnings and the incentive strength across treatments. We define incentive strength as the percentage increase in expected bonus from following the optimal strategy relative to the suboptimal strategy (recall that subjects make a binary decision between batching and not batching). Thresholds were set below expected optimal performance so that earnings were increasing in performance over the relevant range. Per-unit payments were then chosen so that the implied incentive strength was approximately balanced: 14.8\% in \textbf{\textit{IT}}, 15.2\% in \textbf{\textit{GT}} and \textbf{\textit{GT-Nudge}}, and 15.4\% in \textbf{\textit{GT-ST}}. Expected earnings under the optimal strategy were \$4.37, \$4.65, and \$4.22, respectively, ensuring that no treatment was systematically more or less lucrative. Mean realized total payment was \$5.96 (range \$5.32--\$6.81).

For each treatment and strategy pair, we computed expected performance over the experimental horizon using the transient analysis of the corresponding continuous-time Markov chain. In all treatments, the underlying system evolves in continuous time as a finite-state Markov process, with patient arrivals and service completions occurring stochastically. Once a strategy is fixed, the system dynamics are fully characterized by a generator matrix that depends on the treatment-specific assignment rule. For a given generator and performance metric (personal throughput, team throughput, or cumulative sojourn time), we compute the expected performance as follows. We first solve the associated Poisson equation to obtain the long-run average rate of reward and the corresponding relative value function. These objects summarize, respectively, the steady-state performance rate and the incremental future performance associated with ending the experiment in each system state. We then combine these quantities with the transient state distribution at the end of the horizon to obtain the expected cumulative performance starting from an empty and idle system. This approach is standard in the analysis of finite-horizon continuous-time Markov reward processes and allows us to account cleanly for both steady-state behavior and end-of-horizon effects.

To implement the experiment, we require concrete realizations of system evolution. For each treatment, we generate Monte Carlo sample paths from the corresponding continuous-time Markov process under the optimal strategy. We then select two representative paths whose realized incentive strength---both in absolute terms and relative to the optimal benchmark---lies within two percentage points of the theoretical expectation across all treatments. These same sample paths are used consistently across treatments, holding the underlying stochastic realization fixed while varying only the incentive structure.

\subsection{Additional Experimental Results}\label{app:demographics}

\subsubsection*{Supporting Analysis for \textsection 5.1-5.2}
\noindent\textbf{Table~\ref{tab:demographics}.} This table is referenced in \S\ref{subsec:Prolific:Results} and \S\ref{subsec:Physicians}. The table reports the size and demographic composition of the experimental samples, separately by treatment for the Prolific sample and combined for the verified-physician sample. For each subgroup we report the number of participants, the mean and standard deviation of age, the percentage of female participants, and (for the Prolific sample) the distribution of self-reported primary healthcare role.

\begin{table}[!htbp]
\centering\footnotesize
\caption{Participant Demographics by Treatment and Sample}
\label{tab:demographics}
\renewcommand{\arraystretch}{1.15}
\begin{tabular}{lccccc|cc}
\hline
 & \multicolumn{5}{c|}{\textit{Prolific Sample}} & \multicolumn{2}{c}{\textit{Physician Sample}} \\
 & \textbf{\textit{IT}} & \textbf{\textit{GT}} & \textbf{\textit{GT-N}} & \textbf{\textit{GT-ST}} & Overall
 & \textbf{\textit{GT}} & \textbf{\textit{GT-Nudge}} \\
\hline
$N$              & 53   & 51   & 52   & 47   & 203  & 41   & 32 \\
Age, mean (SD)   & 37.7 (13.4) & 39.0 (9.2) & 42.0 (11.6) & 38.1 (11.4) & 39.2 (11.6)
                 & 40.5 (6.9) & 44.2 (10.9) \\
Female (\%)      & 69.8 & 74.5 & 76.9 & 63.8 & 71.4 & 43.9 & 59.4 \\
\hline
\multicolumn{8}{l}{\textit{Primary healthcare role, Prolific (\%): }
Nurse 33.0, Other clinical 18.7, Technician 11.8, Therapist 11.3,} \\
\multicolumn{8}{l}{\quad Physician 8.9, Administrator 8.4, Paramedic 3.9, Pharmacist 3.9.
\textit{ Physician sample: }100\% EM physicians.} \\
\hline
\end{tabular}
\end{table}

\noindent\textbf{Table~\ref{tab:logit_prolific}.} This table is referenced in \S\ref{subsec:Prolific:Results}. This table reports logit regressions in which the dependent variable is an indicator for whether the participant chose the optimal strategy in the Round~2 incentivized round, where the optimal strategy is to batch under \textbf{\textit{IT}} and not to batch under \textbf{\textit{GT}}, \textbf{\textit{GT-Nudge}}, and \textbf{\textit{GT-ST}}. Columns~(1)--(4) include all four treatments with \textbf{\textit{IT}} as the omitted reference category; columns~(5)--(8) restrict the sample to the three group-incentive treatments and use \textbf{\textit{GT}} as the omitted reference. Within each block the four columns add controls stepwise: treatment dummies only, then age and sex, then education fixed effects, then a richer set of professional-experience controls. The nudge significantly increases the probability of choosing the optimal strategy relative to \textbf{\textit{GT}} in every specification ($p<0.05$), while the sojourn-time framing of incentives in \textbf{\textit{GT-ST}} does not significantly differ from \textbf{\textit{GT}}. Demographic variables are not robust predictors of optimal play, although clinical experience predicts optimality among the more experienced participants.

\begin{table}[!htbp]
\centering\scriptsize
\caption{Logit Regressions: Determinants of Optimal Strategy Choice (Prolific Sample)}
\label{tab:logit_prolific}
\renewcommand{\arraystretch}{1.0}
\begin{tabular}{lcccccccc}
\hline
 & \multicolumn{4}{c}{\textbf{\textit{IT, GT, GT-Nudge, GT-ST}}} & \multicolumn{4}{c}{\textbf{\textit{GT, GT-Nudge, GT-ST}}} \\
 & \multicolumn{4}{c}{DV: Chose optimally} & \multicolumn{4}{c}{DV: Chose optimally (did not batch)} \\
 & (1) & (2) & (3) & (4) & (5) & (6) & (7) & (8) \\
\hline
\textbf{\textit{GT}}       & $-5.586^{***}$ & $-5.685^{***}$ & $-5.715^{***}$ & $-6.225^{***}$ & & & & \\
                            & $(0.841)$ & $(0.860)$ & $(0.872)$ & $(0.975)$ & & & & \\[3pt]
\textbf{\textit{GT-Nudge}} & $-4.017^{***}$ & $-4.008^{***}$ & $-4.093^{***}$ & $-4.396^{***}$ & $1.569^{**}$ & $1.667^{**}$ & $1.587^{**}$ & $1.782^{**}$ \\
                            & $(0.679)$ & $(0.692)$ & $(0.718)$ & $(0.789)$ & $(0.680)$ & $(0.688)$ & $(0.697)$ & $(0.731)$ \\[3pt]
\textbf{\textit{GT-ST}}    & $-5.188^{***}$ & $-5.350^{***}$ & $-5.412^{***}$ & $-5.681^{***}$ & $0.398$ & $0.330$ & $0.301$ & $0.505$ \\
                            & $(0.792)$ & $(0.821)$ & $(0.836)$ & $(0.922)$ & $(0.792)$ & $(0.798)$ & $(0.805)$ & $(0.837)$ \\[3pt]
Age                         &  & $-0.035$ & $-0.033$ & $-0.082^{**}$ &  & $-0.030$ & $-0.028$ & $-0.073^{*}$ \\
                            &  & $(0.021)$ & $(0.021)$ & $(0.035)$ &  & $(0.025)$ & $(0.026)$ & $(0.040)$ \\[3pt]
Female                      &  & $0.151$ & $0.179$ & $0.353$ &  & $0.242$ & $0.275$ & $0.462$ \\
                            &  & $(0.526)$ & $(0.526)$ & $(0.559)$ &  & $(0.564)$ & $(0.570)$ & $(0.602)$ \\[3pt]
Direct Care Provider        &  &  &  & $0.801$ &  &  &  & $0.486$ \\
                            &  &  &  & $(0.523)$ &  &  &  & $(0.554)$ \\[3pt]
Patient Assignment Exp.     &  &  &  & $-0.274$ &  &  &  & $-0.397$ \\
                            &  &  &  & $(0.630)$ &  &  &  & $(0.661)$ \\[3pt]
Clinical Experience         &  &  &  & $0.746^{**}$ &  &  &  & $0.676^{*}$ \\
                            &  &  &  & $(0.367)$ &  &  &  & $(0.388)$ \\[3pt]
Income                      &  &  &  & $-0.178$ &  &  &  & $-0.116$ \\
                            &  &  &  & $(0.178)$ &  &  &  & $(0.191)$ \\[3pt]
Education FE                &  &  & Yes & Yes &  &  & Yes & Yes \\
\hline
$N$                         & 203 & 203 & 203 & 203 & 150 & 150 & 150 & 150 \\
$R^2$                       & 0.503 & 0.514 & 0.517 & 0.550 & 0.067 & 0.084 & 0.100 & 0.143 \\
\hline
\multicolumn{9}{l}{\textit{Post-hoc Wald tests ($p$-values)}} \\
\hline
\textbf{\textit{GT}} $=$ \textbf{\textit{GT-Nudge}}    & $0.021$ & $0.015$ & $0.020$ & $0.012$ & $0.021$ & $0.015$ & $0.023$ & $0.015$ \\
\textbf{\textit{GT}} $=$ \textbf{\textit{GT-ST}}        & $0.616$ & $0.674$ & $0.706$ & $0.517$ & $0.616$ & $0.679$ & $0.709$ & $0.546$ \\
\textbf{\textit{GT-Nudge}} $=$ \textbf{\textit{GT-ST}}  & $0.058$ & $0.034$ & $0.039$ & $0.056$ & $0.058$ & $0.036$ & $0.047$ & $0.061$ \\
\hline
\end{tabular}
\\[4pt]
\begin{minipage}{0.88\textwidth}
\scriptsize
\textit{Notes.} $^{*}p<0.10$, $^{**}p<0.05$, $^{***}p<0.01$. Standard errors in parentheses. DV $= 1$ if participant chose the optimal strategy. Columns (1)--(4): all four treatments, reference $=$ \textbf{\textit{IT}}. Columns (5)--(8): \textbf{\textit{GT}}, \textbf{\textit{GT-Nudge}}, \textbf{\textit{GT-ST}} only; reference $=$ \textbf{\textit{GT}}. Direct Care Provider $= 1$ if physician, nurse, or paramedic. Patient Assignment Exp.\ $= 1$ if participant reports assigning patients in their role. Clinical Experience is self-reported on a 0--5 scale. Post-hoc tests report $p$-values from Wald tests of coefficient equality.
\end{minipage}
\end{table}

\noindent\textbf{Table~\ref{tab:round1_regret}.} This table is referenced in \S\ref{subsec:Prolific:Results}. It reports decision-level logit regressions on the Round~1 (practice round) data, with the dependent variable equal to one if the participant batched at decision $t$. The variable of interest, ``case still there at $t$,'' equals one if the participant chose not to batch at $t-1$ when batching was feasible and the unassigned count at $t$ is at least as large as at $t-1$, so the case the participant declined to claim was plausibly still in the queue. Columns~(1) and~(2) use the full sample of decisions where batching is feasible at $t$ ($N=2{,}109$ decisions, 203 participants), with column~(2) additionally controlling for batching at $t-1$; column~(3) restricts to decisions where the participant chose not to batch at $t-1$ when batching was feasible ($N=786$ decisions, 148 participants). Across columns~(2) and~(3), participants are about 10 percentage points more likely to batch at $t$ when the case left behind was still in the queue ($p$-values between 0.013 and 0.018). Note that the time-elapsed coefficient is positive overall (col.~1) but negative in col.~(3) because, as the round progresses, the non-batcher subsample becomes dominated by participants who have repeatedly declined to batch and are unlikely to switch regardless of the feedback signal.

\begin{table}[!htbp]
\centering\scriptsize
\caption{Round 1 Decision-Level Regression: Effect of ``Case Still There'' on Next-Decision Batching}
\label{tab:round1_regret}
\renewcommand{\arraystretch}{1.0}
\begin{tabular}{lccc}
\hline
 & \multicolumn{3}{c}{DV: Batch at decision $t$} \\
 & (1) & (2) & (3) \\
\hline
Case still there at $t$  &                & $0.631^{**}$   & $0.627^{**}$ \\
                         &                & $(0.262)$      & $(0.266)$ \\[3pt]
Batched at $t{-}1$       &                & $3.679^{***}$  &              \\
                         &                & $(0.324)$      &              \\[3pt]
Time elapsed             & $0.068^{***}$  & $-0.007$       & $-0.056^{**}$ \\
                         & $(0.017)$      & $(0.016)$      & $(0.028)$ \\[3pt]
(Intercept)              & $0.164$        & $-1.653^{***}$ & $-1.789^{***}$ \\
                         & $(0.258)$      & $(0.312)$      & $(0.346)$ \\
\hline
Treatment FE             & Yes            & Yes            & Yes \\
Sample                   & Full           & Full           & Non-batchers \\
\hline
$N$ (decisions)          & 2{,}109        & 2{,}109        & 786 \\
$N$ (participants)       & 203            & 203            & 148 \\
\hline
\end{tabular}
\\[2pt]
\begin{minipage}{0.45\textwidth}
\scriptsize
\textit{Notes.} $^{*}p<0.10$, $^{**}p<0.05$, $^{***}p<0.01$. Logit, cluster-robust SE at the participant level. DV $=1$ if the participant batches at decision $t$. \textit{Case still there at $t$} $=1$ if the participant chose not to batch at $t-1$ (with $\text{prev\_U}\geq 2$) and the unassigned count at $t$ is $\geq \text{prev\_U}$. \textit{Time elapsed} in tens of model-time units (round $=600$). Cols (1)--(2): full Round~1 sample. Col (3): subsample with $\text{prev\_choice}=1$ and $\text{prev\_U}\geq 2$.
\end{minipage}
\end{table}

\subsubsection*{Supporting Analysis for \textsection 5.3\label{app:quote_analysis}}

This subsection describes the procedure used to classify participants' open-text responses to the post-experiment question \emph{``Please describe the strategy you chose and the reasoning behind it''} (referenced in \textsection5.3). The procedure is a keyword-based classification of each response into one of three motives, with responses that match none of the three sets of keywords left uncoded.

\textbf{Classification scheme.} We classify each response as articulating one of three modes of thinking:
\begin{itemize}
\item \textit{Busyness}: the participant articulates a preference for being busy or avoiding idle time. Keywords include ``busy'', ``idle'', ``waste time'', ``back and forth'', ``lazy'', ``multitask'', ``faster pace'', ``more productive'', ``save time'', and ``lag time''.
\item \textit{Throughput}: the participant articulates a belief that batching maximizes the number of patients treated. Keywords include ``more patients'', ``as many patients as possible'', ``maximize patients/treated'', ``treat more'', ``clear queue'', ``move quickly'', ``free up space'', ``benefit for the group''.
\item \textit{Habit}: the participant invokes their real-world practice without articulating a busyness or throughput reason. Keywords include ``in real life'', ``what I do at work'', ``what I usually do'', ``I am used to'', ``based on how I practice'', and ``in all my years''.
\end{itemize}
The classification has a strict precedence (\textit{habit} before \textit{busyness} before \textit{throughput}) for responses that match keywords in more than one set, so that habit-language responses are not coded as busyness even when they include words like ``easier'' or ``efficient''. A small share of non-batchers articulate the system-optimal logic (partner-idle reasoning) and are coded as ``other''. Table~\ref{tab:quote_classifications} reports the share of batchers and non-batchers in each treatment whose response was classified into each category, separately for the Prolific and physician samples.

\begin{table}[!htbp]
\centering\scriptsize
\caption{Classification of Open-text Responses.}
\label{tab:quote_classifications}
\renewcommand{\arraystretch}{1.1}
\setlength{\tabcolsep}{4pt}
\begin{tabular}{l|ccccc|ccc}
\hline
 & \multicolumn{5}{c|}{\textit{Prolific}} & \multicolumn{3}{c}{\textit{Physician}} \\
 & \textbf{\textit{IT}} & \textbf{\textit{GT}} & \textbf{\textit{GT-Nudge}} & \textbf{\textit{GT-ST}} & Pooled & \textbf{\textit{GT}} & \textbf{\textit{GT-Nudge}} & Pooled \\
\hline
\multicolumn{9}{l}{\textit{Panel A: Batchers}} \\
\hline
N batchers (with description) & 50 & 48 & 40 & 43 & 181 & 21 & 11 & 32 \\
Busyness ($\%$)           & 10.0 & 12.5 & 10.0 & 25.6 & 14.4 & 38.1 & 36.4 & 37.5 \\
Throughput ($\%$)         & 50.0 & 64.6 & 67.5 & 46.5 & 56.9 & 33.3 & 27.3 & 31.2 \\
Habit ($\%$)              & 2.0  & 0.0  & 0.0  & 7.0  & 2.2  & 9.5  & 27.3 & 15.6 \\
Other ($\%$)              & 38.0 & 22.9 & 22.5 & 20.9 & 26.5 & 19.0 & 9.1  & 15.6 \\
\hline
\multicolumn{9}{l}{\textit{Panel B: Non-batchers}} \\
\hline
N non-batchers (with description) & 3 & 3 & 12 & 4 & 22 & 7 & 8 & 15 \\
Busyness ($\%$)           & 0.0   & 0.0   & 0.0  & 25.0 & 4.5  & 14.3 & 12.5 & 13.3 \\
Throughput ($\%$)         & 0.0   & 0.0   & 16.7 & 0.0  & 9.1  & 14.3 & 0.0  & 6.7 \\
Habit ($\%$)              & 0.0   & 0.0   & 0.0  & 0.0  & 0.0  & 0.0  & 0.0  & 0.0 \\
Other ($\%$)              & 100.0 & 100.0 & 83.3 & 75.0 & 86.4 & 71.4 & 87.5 & 80.0 \\
\hline
\multicolumn{9}{l}{\textit{Panel C: All participants (batchers $+$ non-batchers)}} \\
\hline
N (with description)      & 53   & 51   & 52   & 47   & 203  & 28   & 19   & 47 \\
Busyness ($\%$)           & 9.4  & 11.8 & 7.7  & 25.5 & 13.3 & 32.1 & 26.3 & 29.8 \\
Throughput ($\%$)         & 47.2 & 60.8 & 55.8 & 42.6 & 51.7 & 28.6 & 15.8 & 23.4 \\
Habit ($\%$)              & 1.9  & 0.0  & 0.0  & 6.4  & 2.0  & 7.1  & 15.8 & 10.6 \\
Other ($\%$)              & 41.5 & 27.5 & 36.5 & 25.5 & 33.0 & 32.1 & 42.1 & 36.2 \\
\hline
\end{tabular}
\begin{minipage}{0.75\textwidth}
\smallskip
\scriptsize
\textit{Notes.} The \textit{other} category collects responses coded as system-optimal (correct partner-idle reasoning, mostly given by non-batchers), turn-away concern, team consideration, incentive response, and responses too short or vague to classify.
\end{minipage}
\end{table}

\subsection{Experiment Instructions}\label{app:instructions}

This section reproduces the experiment instructions shown to participants. The Prolific version (which includes both the practice round and the strategy-commitment round) is shown first, followed by the physician version.

\subsubsection{Prolific Version: Instructions}

The following instructions were displayed to all Prolific participants. The highlighted earnings paragraph varied by treatment; all four variants are shown.

\medskip
\begin{mdframed}[innerleftmargin=10pt, innerrightmargin=10pt, innertopmargin=6pt, innerbottommargin=6pt, linewidth=0.4pt]
\footnotesize
In this study you will be asked to play the role of a physician treating patients in an emergency department.

Patients arrive in groups of 3 throughout your shift. There are two treating physicians: you and your partner. Together, you can have up to 4 patients in the system at any time (waiting or being treated). Your task is to self-assign patients from the unassigned pool and treat them.

\textbf{How the system works:}
\begin{itemize}
\item \textbf{Shift Duration:} Your shift lasts exactly 600 time units.
\item \textbf{Patient Intake:} Patient groups arrive at random times, with an average of 30 time units between arrivals.
\item \textbf{Treatment Times:} Each patient requires a different amount of treatment time, averaging 15 time units per patient.
\item \textbf{Sequential Treatment:} When you self-assign multiple patients, you treat them one at a time in sequence (you complete the first patient before starting the second). From the patient's perspective, it makes no difference whether they wait in the waiting room or in a treatment room.
\item \textbf{Partner Behavior:} Your partner's decisions are pre-programmed. In particular, your partner is programmed to always self-assign exactly one patient. Your partner's treatment times are the same as yours. When both you and your partner are idle and a group of patients arrives, your partner assigns first, then you make your decision.
\end{itemize}

\textbf{Your decisions:} Each time there are two or more unassigned patients waiting, you will be asked to make a decision. At each decision point, you can either:
\begin{enumerate}
\item Self-assign \textbf{one} patient and begin treatment, OR
\item Self-assign \textbf{two} patients and begin treatment.
\end{enumerate}

\textbf{Earnings variants by treatment:}

\textit{IT (Individual Throughput):} Your earnings are based on the number of patients treated by you. After you complete your first 6 patients, you will begin earning \$0.24 for each additional patient you successfully treat. This is in addition to the participation fee of \$2.00.

\textit{GT / GT-Nudge (Group Throughput):} Your earnings are based on the number of patients treated by both you and your partner combined. After you and your partner together complete 36 patients, you will begin earning \$0.60 for each additional patient treated by either you or your partner. This is in addition to the participation fee of \$2.00.

\textit{GT-ST (Sojourn Time):} Your earnings are based on the time patients spend in the system. There is a target of 1{,}200 time units of total time spent by patients (waiting and being treated by both you and your partner combined). For every time unit below this target, you will earn \$0.014 (1.4 cents). This is in addition to the participation fee of \$2.00.

\smallskip

\textbf{How patient arrivals work:} As noted earlier, patients always arrive in groups of 3. However, the number of patients that can be taken in and actually enter the system depends on how many patients are already in the system (waiting or being treated) when they arrive. The system has a capacity of 4 patients.

\begin{itemize}
\item \textbf{0 or 1 patients currently in the system:} All 3 new patients are taken in.
\item \textbf{2 patients currently in the system:} 2 new patients are taken in, 1 is turned away.
\item \textbf{3 patients currently in the system:} 1 new patient is taken in, 2 are turned away.
\item \textbf{4 patients currently in the system:} No new patients are taken in. All 3 are turned away.
\end{itemize}

\textit{Turned away patients are diverted to another facility and do not return. They cannot be treated by you or your partner.}

\smallskip

\textbf{Variability in Times:} There will be variability in how long each patient occupies a treatment room. As such, some patients will stay shorter than 15 time units and some will stay longer than 15 time units. Similarly, the time between patient arrivals will vary; some intervals will be shorter than 30 time units and some will be longer. [Participants could expand collapsible details to view the underlying treatment-time and inter-arrival-time distributions, both exponential.]

\smallskip

\textbf{End-of-shift procedures:}

\textit{Shift Termination:} Your shift ends exactly at 600 time units. At that point, you will stop treating patients immediately.

\textit{Final Payment Adjustment.} \textit{IT:} Your payment will be based on the number of patients you treat during the shift. However, depending on the state of the system when the shift ends, your compensation may be adjusted. \textit{GT / GT-Nudge:} Your payment will be based on the number of patients you and your partner treat during the shift. However, depending on the state of the system when the shift ends, your compensation may be adjusted. \textit{GT-ST:} Your payment is based on the time spent by all the patients treated by you and your partner during the shift. However, if there are patients still in the system when the shift ends (waiting or being treated), their expected remaining time will be added to your total time spent, reducing your time-saved earnings.

\textit{Why this matters.} \textit{IT / GT / GT-Nudge:} The emergency department needs to ensure continuity of care. Your decisions about how many to take at once affect not just how many patients you complete during the shift, but also the state of the system when the shift ends. The payment adjustment ensures that your compensation fairly reflects both your productivity during the shift and the condition you leave the department in. \textit{GT-ST:} The emergency department needs to ensure continuity of care. Your decisions about how many to take at once affect not just the total time patients spend in the system during your shift, but also how many patients remain when the shift ends. The payment adjustment ensures that your compensation fairly reflects the total time burden on patients, including those left in the system.

\textit{Adjustment Table.} Each participant saw a table corresponding to their treatment. The three variants are shown below.

\smallskip
\renewcommand{\arraystretch}{1.05}

\textit{IT (Individual Throughput):}\\[2pt]
\centering
\begin{tabular}{>{\centering\arraybackslash}p{2.0cm}>{\centering\arraybackslash}p{2.0cm}>{\centering\arraybackslash}p{2.0cm}>{\centering\arraybackslash}p{3.6cm}}
\hline
Unassigned & Assigned to You & Assigned to Partner & Additional Patients Credited \\
\hline
0 & 0 & 0 & 0.000 \\
0 & 0 & 1 & 0.038 \\
0 & 1 & 0 & 0.705 \\
0 & 2 & 0 & 1.174 \\
0 & 1 & 1 & 0.720 \\
0 & 2 & 1 & 1.181 \\
1 & 1 & 1 & 0.999 \\
1 & 2 & 1 & 1.294 \\
2 & 1 & 1 & 1.294 \\
\hline
\end{tabular}

\smallskip
\textit{GT / GT-Nudge (Group Throughput):}\\[2pt]
\centering
\begin{tabular}{>{\centering\arraybackslash}p{2.0cm}>{\centering\arraybackslash}p{2.0cm}>{\centering\arraybackslash}p{2.0cm}>{\centering\arraybackslash}p{3.6cm}}
\hline
Unassigned & Assigned to You & Assigned to Partner & Additional Patients Credited \\
\hline
0 & 0 & 0 & 0.00 \\
0 & 0 & 1 & 0.82 \\
0 & 1 & 0 & 0.82 \\
0 & 2 & 0 & 1.36 \\
0 & 1 & 1 & 1.55 \\
0 & 2 & 1 & 2.05 \\
1 & 1 & 1 & 2.13 \\
1 & 2 & 1 & 2.56 \\
2 & 1 & 1 & 2.60 \\
\hline
\end{tabular}

\smallskip
\textit{GT-ST (Sojourn Time):}\\[2pt]
\centering
\begin{tabular}{>{\centering\arraybackslash}p{2.0cm}>{\centering\arraybackslash}p{2.0cm}>{\centering\arraybackslash}p{2.0cm}>{\centering\arraybackslash}p{3.6cm}}
\hline
Unassigned & Assigned to You & Assigned to Partner & Time Units Added \\
\hline
0 & 0 & 0 & 0.00 \\
0 & 0 & 1 & 2.59 \\
0 & 1 & 0 & 2.59 \\
0 & 2 & 0 & 5.99 \\
0 & 1 & 1 & 4.63 \\
0 & 2 & 1 & 7.80 \\
1 & 1 & 1 & 7.26 \\
1 & 2 & 1 & 10.58 \\
2 & 1 & 1 & 10.58 \\
\hline
\end{tabular}

\end{mdframed}
\medskip

\subsubsection{Prolific Version: Strategy Selection (Round 2)}

After the practice round, participants were presented with the following strategy-selection screen. The screen below shows the \textbf{\textit{GT}} and \textbf{\textit{GT-Nudge}} variant; the screens for the other treatments are analogous, with the appropriate earnings reminder substituted in the place indicated.

\medskip
\begin{mdframed}[innerleftmargin=10pt, innerrightmargin=10pt, innertopmargin=6pt, innerbottommargin=6pt, linewidth=0.4pt]
\footnotesize
\textbf{Take a moment to reflect:} Now that you were able to experiment with different strategies and see how the system operates, please think carefully about which approach will maximize your earnings. \textit{[Earnings reminder varied by treatment.]}

\textit{[GT-Nudge treatment only:]} \textbf{Note:} If you choose Strategy 2 (Self-assign 2 patients whenever possible), the following can occur: While you are treating your two patients, your partner becomes idle and is not treating any patients.

Select a strategy:
\begin{itemize}
\item \textbf{Strategy 1:} Always self-assign 1 patient
\item \textbf{Strategy 2:} Always self-assign 2 patients (whenever possible)
\end{itemize}
\end{mdframed}
\medskip

\subsubsection{Physician Version}

The physician version of the experiment was designed to minimize the time burden on practicing emergency physicians. Three changes were made relative to the Prolific version. First, the practice round (Round 1) was omitted; physicians read the instructions and proceeded directly to the strategy-commitment decision. Second, the comprehension quiz was removed. Third, the instruction text was condensed into a shorter, bullet-point format covering the same information (system parameters, partner behavior, capacity constraints, and earnings structure). The earnings structure and nudge message were identical to the Prolific \textbf{\textit{GT}} and \textbf{\textit{GT-Nudge}} treatments. These modifications reduced mean completion time from 15.2 minutes (Prolific) to 7.2 minutes (physicians).

\subsubsection{Comprehension Quiz (Prolific Only)}

Prolific participants answered the following comprehension questions (with feedback and retry on incorrect answers):

\medskip
\begin{mdframed}[innerleftmargin=10pt, innerrightmargin=10pt, innertopmargin=6pt, innerbottommargin=6pt, linewidth=0.4pt]
\footnotesize
\noindent\textbf{Q1:} I have to treat all patients; True or False?\\
(a) True \quad (b) False, my partner will treat all the patients. \quad \textbf{(c) False, my partner and I will together treat the patients.}
\medskip

\noindent\textbf{Q2:} The task is entirely hypothetical and I will receive a fixed payment at the end of the experiment; True or False?\\
\textit{[Answer options varied by treatment to reflect the specific incentive structure.]}
\medskip

\noindent\textbf{Q3:} Are your partner's decisions computerized? If so, what does your partner do?\\
(a) No, the partner is a real human. \quad (b) Yes, partner randomly decides. \quad \textbf{(c) Yes, partner always self-assigns one patient.} \quad (d) Yes, partner always self-assigns two patients.
\medskip

\noindent\textbf{Q4:} All the patients are identical and all arrive at the same time. True or False?\\
(a) True \quad (b) False, the patients arrive at different points in time. \quad (c) False, the patients have different treatment times. \quad \textbf{(d) False, the patients arrive at different times AND have different treatment times.}
\medskip

\noindent\textbf{Q5:} When a batch of patients arrives, how many are admitted to the system?\\
(a) Always 3 patients. \quad \textbf{(b) 0, 1, 2, or 3 patients, depending on how many patients are already in the system.} \quad (c) 1 or 2 patients, depending on my choice. \quad (d) 0 patients if either physician is busy.
\end{mdframed}
\medskip

\subsection{Additional Experiment Interface Screenshots}\label{app:screenshots}

\begin{figure}[!t]
\centering
\caption{Strategy choice screen (\textbf{\textit{GT-Nudge}} treatment), with the nudge message highlighted.}
\label{fig:exp_strategy_nudge}
\includegraphics[width=0.85\textwidth, trim=0 0 0 38, clip]{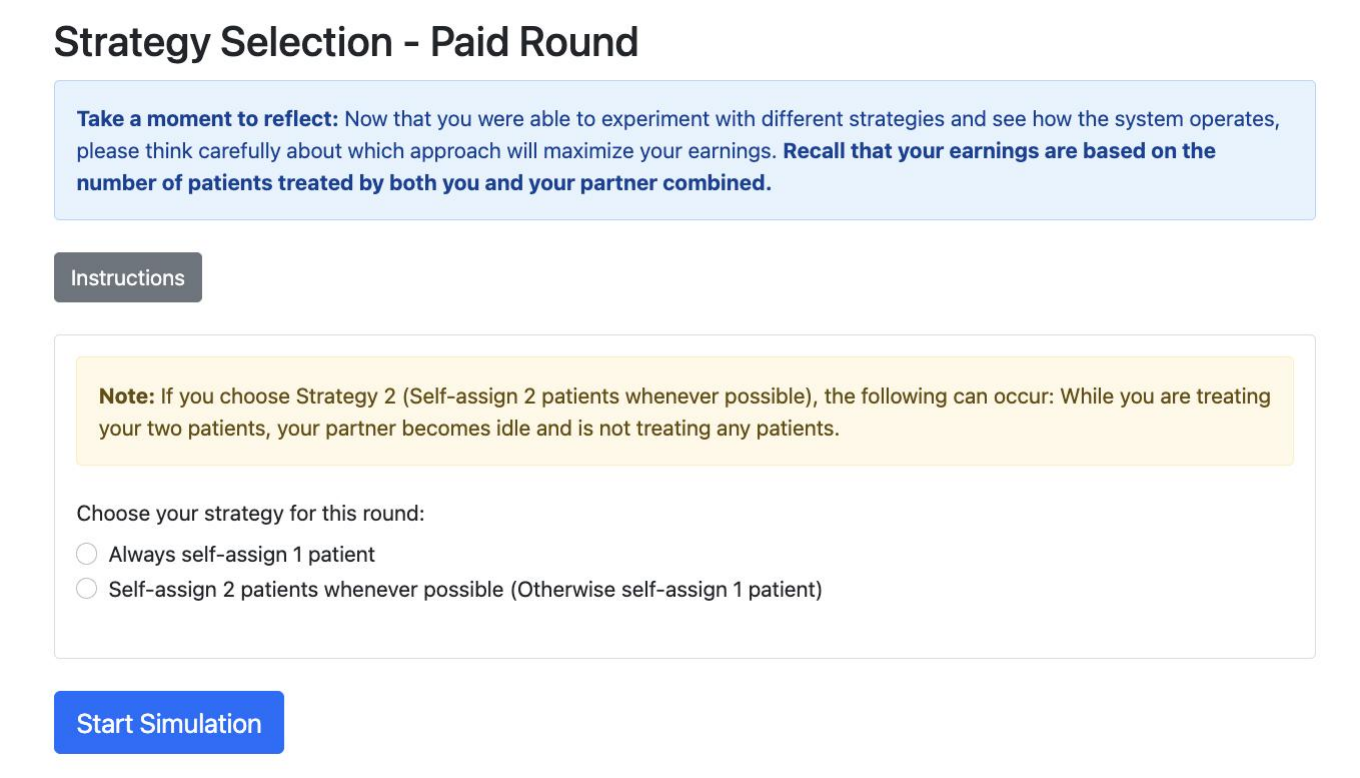}
\end{figure}

\begin{figure}[!t]
\centering
\caption{Round 2 simulation: the committed strategy executes automatically over the full shift.}
\label{fig:exp_simulation}
\includegraphics[width=0.8\textwidth, trim=0 0 0 70, clip]{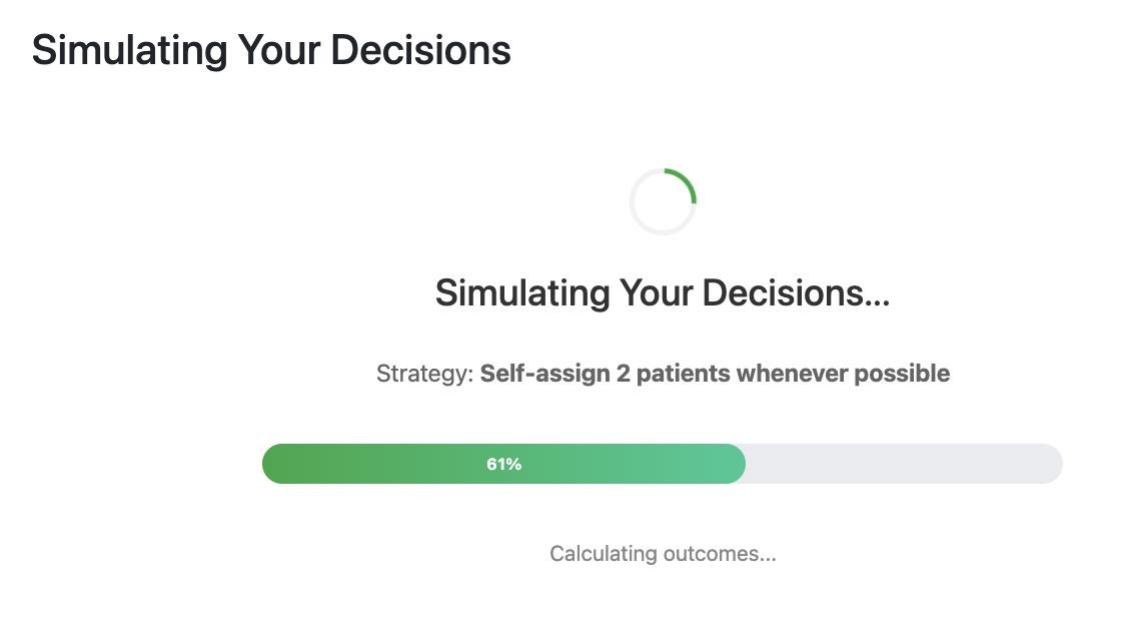}
\end{figure}

This appendix supplements Figure~\ref{fig:exp_interface} in the main text by showing the remaining screens of the experimental interface. Figure~\ref{fig:exp_strategy_nudge} shows the Round 2 strategy-commitment screen in the \textbf{\textit{GT-Nudge}} treatment, the variant of panel (c) of Figure~\ref{fig:exp_interface} with the nudge message highlighted. Figure~\ref{fig:exp_simulation} shows the simulation screen that runs after the participant commits to a strategy.

\vspace{-4cm}

\end{document}